\documentclass[a4paper,12pt]{article}
\pdfoutput=1
\usepackage{latexsym}
\usepackage{amsfonts}
\usepackage{amsmath}
\usepackage{amsthm}
\usepackage{unicode}
\usepackage[english]{babel}
\usepackage[utf8]{inputenc}
\usepackage[T1]{fontenc}
\usepackage[normalem]{ulem}
\usepackage{array}
\usepackage{amssymb}
\usepackage{graphicx}
\usepackage{url}
\usepackage{float}
\usepackage{subfloat}
\usepackage{bm}
\usepackage[
style=apa,
backend=biber,
style=numeric,
citestyle=nature,
sorting=nyt, 
maxcitenames=2, 
giveninits=true,
isbn=false,
doi=false,
url=false,
natbib=true,
]{biblatex}
\DeclareFieldFormat[article,periodical]{volume}{\mkbibbold{#1}}
\DeclareCaseLangs{}
\usepackage{csquotes}
\usepackage{caption}
\usepackage{lscape} 
\usepackage{afterpage}
\usepackage{color} \usepackage{floatpag}
\floatpagestyle{empty}

\usepackage[export]{adjustbox}

\def\ind{\mathrel{\hbox{\rlap{\hbox to 7.5pt{\hrulefill}}\raise6.6pt\hbox{\eka\char'167}}}}

\newtheorem{lemma}{Lemma}[section]

\numberwithin{equation}{section}

\def\bftau{\boldsymbol{\tau}}

\usepackage{booktabs,array}

\newcount\rowc

\makeatletter
\def\ttabular{%
\hbox\bgroup
\let\\\cr
\def\rulea{\ifnum\rowc=\@ne \hrule height 1.3pt \fi}
\def\ruleb{
\ifnum\rowc=1\hrule height 1.3pt \else
\ifnum\rowc=6\hrule height \heavyrulewidth 
   \else \hrule height \lightrulewidth\fi\fi}
\valign\bgroup
\global\rowc\@ne
\rulea
\hbox to 10em{\strut \hfill##\hfill}%
\ruleb
&&%
\global\advance\rowc\@ne
\hbox to 10em{\strut\hfill##\hfill}%
\ruleb
\cr}
\def\endttabular{\crcr\egroup\egroup}

\usepackage[ruled,vlined]{algorithm2e}
\usepackage[noend]{algpseudocode}

\usepackage{subcaption}
\usepackage{wrapfig}
\usepackage{wasysym}
\usepackage{enumitem}
\usepackage{adjustbox}
\usepackage{ragged2e}
\usepackage[svgnames,table]{xcolor}
\usepackage{tikz}
\usepackage{longtable}
\usepackage{changepage}
\usepackage{setspace}
\usepackage{hhline}
\usepackage{multicol}
\usepackage{tabto}
\usepackage{float}
\usepackage{multirow}
\usepackage{makecell}
\usepackage{fancyhdr}
\usepackage[toc,page]{appendix}
\usepackage{authblk}
\usetikzlibrary{shapes.symbols,shapes.geometric,shadows,arrows.meta}
\tikzset{>={Latex[width=1.5mm,length=2mm]}}
\usepackage{flowchart}\usepackage[paperheight=11.69in,paperwidth=8.27in,left=0.79in,right=0.79in,top=0.98in,bottom=0.98in,headheight=1in]{geometry}
\TabPositions{0.91in,1.82in,2.73in,3.64in,4.55in,5.46in,6.37in,}
\usepackage{xr-hyper} 
\usepackage{hyperref}
\hypersetup{
colorlinks=false,
}
\makeatletter
\newcommand*{\addFileDependency}[1]{
  \typeout{(#1)}
  \@addtofilelist{#1}
  \IfFileExists{#1}{}{\typeout{No file #1.}}
}

\makeatother

\setlistdepth{9}
\renewlist{enumerate}{enumerate}{9}
		\setlist[enumerate,1]{label=\arabic*)}
		\setlist[enumerate,2]{label=\alph*)}
		\setlist[enumerate,3]{label=(\roman*)}
		\setlist[enumerate,4]{label=(\arabic*)}
		\setlist[enumerate,5]{label=(\Alph*)}
		\setlist[enumerate,6]{label=(\Roman*)}
		\setlist[enumerate,7]{label=\arabic*}
		\setlist[enumerate,8]{label=\alph*}
		\setlist[enumerate,9]{label=\roman*}

\renewlist{itemize}{itemize}{9}
		\setlist[itemize]{label=$\cdot$}
		\setlist[itemize,1]{label=\textbullet}
		\setlist[itemize,2]{label=$\circ$}
		\setlist[itemize,3]{label=$\ast$}
		\setlist[itemize,4]{label=$\dagger$}
		\setlist[itemize,5]{label=$\triangleright$}
		\setlist[itemize,6]{label=$\bigstar$}
		\setlist[itemize,7]{label=$\blacklozenge$}
		\setlist[itemize,8]{label=$\prime$}
\def\R{\mathbb{R}}

\def\N{\mathbb{N}}
\def\E{\mathbb{E}}

\def\P{\mathbb{P}}

\DeclareBoldMathCommand\bfalpha{\alpha}
\DeclareBoldMathCommand\bftheta{\theta}
\DeclareBoldMathCommand\bfrho{\rho}
\DeclareBoldMathCommand\bfeta{\eta}

\title{Is control of  type I error rate needed in Bayesian clinical trial designs?}

\author[1]{Elja Arjas
\thanks{corresponding author, elja.arjas@helsinki.fi}
}

\author[2]{Dario Gasbarra}
\affil[1]{University of 
Helsinki,
Dept. of Mathematics
and Statistics, P.O. Box 68  
FI-00014 HELSINGIN YLIOPISTO,  Finland
}
\affil[2]{University of Vaasa, Dept. of Mathematics and Statistics, P.O. Box 700 FI-65101 Vaasa Finland}

\date{\today}
\begin{document}
\maketitle

\begin{abstract}

Practical employment of Bayesian trial designs is still rare. Even if accepted in principle, the regulators have commonly required that such designs be calibrated according to an upper bound for the frequentist type I error rate.  This represents an internally inconsistent hybrid methodology, where important advantages from following the Bayesian principles are lost. In practice, in sequential designs, all pre-planned interim looks need to be accounted for in computing type I error rate. If a fixed upper bound on that error rate is imposed, interim looks lower the power of detecting a true effect.

To present an alternative approach, we consider the prototype case of a 2-arm superiority trial with dichotomous outcomes. The design is adaptive,  using error control based on sequentially updated posterior probabilities, to conclude efficacy of the experimental treatment or futility of the trial.  As gatekeepers for a proposed design, the regulators have the main responsibility in determining acceptable levels for the  parameters of the control of false positives, whereas the trial sponsors and investigators will have a natural role in specifying the criteria for stopping the trial due to futility. It is suggested that the traditional frequentist operating characteristics in the design, type I  and type II error rates, be replaced, respectively, by Bayesian criteria called 
\textit{False Discovery Probability} (FDP) and \textit{False Futility Probability} (FFP); both terms correspond directly to their probability interpretations. 
Importantly, sequential error control during the data analysis based on posterior probabilities will satisfy these numerical criteria automatically, without need of preliminary computations before the trial is started.   
The method contains the option of applying a decision rule for terminating the trial, without concluding either efficacy or futility, if the predicted costs from continuing would exceed the corresponding gains.
 
\end{abstract}



\section{Introduction} 

A key issue in the literature on adaptive trial designs is the choice of the statistical paradigm: frequentist, Bayesian, and sometimes a hybrid of these. This choice is not only of a theoretical character but has important practical consequences as well. In particular, in the frequentist approach all pre-planned interim looks have an inflating multiplicity effect on type I error rate. If this error rate is bounded from above by a selected $\alpha$-level, such looks lower the power of the tests that are used. In the Bayesian approach, in contrast, these considerations are redundant because of its reliance on the likelihood principle (see, e.g., \cite{berger}, \cite{doi:10.1198/000313006X117837}). The design may then in principle allow even for optional interim looks into the accumulated outcome data if such decisions do not depend on the model parameters and thereby do not contribute to the likelihood expression. Following the likelihood principle then means that conclusions from the statistical data analysis are the same as what would be obtained had the timing of the interim analyses been included in the design. The result is a much greater flexibility and freedom in selecting and applying adaptive decision rules for running the trial. 

Another reason, not less important in its support of the Bayesian methodology, is that posterior probabilities based on trial data provide direct answers to the questions that are of interest, together with a quantification of the uncertainties that are then involved. For example, one may be led to a conclusion of the following kind: "Based on expert judgment, empirical findings from comparable earlier trials and from the observed trial data, there is at least ninety-five percent probability that the tested experimental treatment is more efficacious, by the margin of the pre-specified  MID, than the standard treatment that was used as control." This is in contrast to conventional Null Hypothesis Significance Testing (NHST) framework, where a positive result from the trial, such as concluded efficacy of a new drug, is registered indirectly, as rejection of the Null. The considered test statistic has a probability interpretation only when viewed as the quantile of the sampling distribution based on the  fixed parameter value corresponding to the Null. For example, in a $t$-test, on observing the value of the test statistic, one first determines the  tail area under the Null, and only then compares this value  to the selected significance level.  

A more elaborate approach to modeling based on the same principles allows also utilization of the tools from Bayesian decision theory (e.g., \cite{robert2007bayesian}), in which the concrete consequences of the possible conclusions from the trial are assessed and then expressed in terms of a utility function. In particular, one could consider the monetary gains from correct and the losses from incorrect conclusions.   

The arguments in support of the Bayesian methodology in the context of clinical trials have been clear for decades, at least since \cite{berry1985interim} and \cite{SPIEGELHALTER19868}. However, progress in their practical application has been slow; for reasons, see e.g., \cite{chevret2012bayesian} and \cite{Lee2012BayesianCT}. Recent accounts of Bayesian methodology in clinical trials can be found in \cite{lin2020novel}, \cite{ijerph18020530}, \cite{zhou_ji_2023} and \cite{muehlemann2023tutorial}. Worth reading is also \cite{ruberg2023application}, presenting views on the current position of Bayesian methods in drug development. For a systematic study of the area, the monographs \cite{berry} and \cite{lesaffre2020} are recommended.

Every clinical trial design involving human subjects. if it is to be implemented in practice, needs to be approved by the regulatory authorities responsible. Even if the statistical analysis of the trial data would be done in accordance with Bayesian principles, the regulators, as exemplified by the  \cite{fda_71512}, \cite{fda_78495} and  \cite{EMA_CHMP_44762_2017} guidelines, have commonly required that the consequent type I error rate remains below a given significance level. Strong logical arguments in support of the conditionality principle, as presented, e.g., in  \cite{berger1988statistical}, have not shaken this nearly axiomatic status of type I error control in trial designs. In consequence, there is currently a large body of literature on how the parameter values of a Bayesian design might be adjusted to satisfy such a requirement. Examples include \cite{freedman1994and}, \cite{ventz2015bayesian}, \cite{nikolakopoulos2016hybrid}, \cite{doi:10.1177/0962280215595058},
\cite{ventz2017combining}, \cite{kopp2019monitoring}, \cite{yu2019unified} and \cite{shi2019control}. Notwithstanding the high technical level of these contributions, such strict numerical control of type I error rate generally dampens the advantages of Bayesian designs. The result is a conceptual and methodological hybrid, with elements taken from two different, mutually inconsistent statistical paradigms.
Due to our general opposition against such hybrid methodology, we do not review 
this literature in more detail; however, if desired, \cite{yu2019unified} is useful reading for such a purpose. 

Instead, and following many earlier authors, we propose that the control of false positives be performed by direct bounding, with a selected error tolerance, of the corresponding posterior probabilities during the data analysis. With this, such error control is not affected by multiplicity considerations in   sequentially determined interim analyses. 

On the other hand, the regulators' acceptance of a design must obviously be based on an argument that is made explicit already before the trial is started.
For this purpose, to replace type I error rate as an operating characteristic, we propose an alternative Bayesian criterion called \textit{False Discovery Probability}
(FDP). It has the required natural probability interpretation in the context, as   the conditional probability that the conclusion is in fact false, given that a positive conclusion from the trial is established. The criterion is directly connected, via a conditional expectation, to the corresponding posterior probabilities 
that will arise when the trial is run. 

While the regulators' main interest and responsibility lies naturally in the control of false positives, the sponsors may be more concerned with limiting the possibility of false negatives. In traditional designs, this  amounts to calculating in advance a sample size large enough to  guarantee, at a given significance level and a given effect difference, a high enough power or, equivalently, small type II error rate. In  Bayesian trial designs such preliminary sample size calculations, and their Bayesian variants (e.g., \cite{10.1214/088342304000000080}, \cite{lee2024using}), become largely redundant, as these designs  allow in principle for 
continued reassessment of the results, without need to thereby adjust the error bounds.

Similarly to the FDP criterion replacing type I error rate, we propose that a criterion called  \textit{False Futility Probability} (FFP) would replace type II error rate as an operating characteristic of the design. It, too, arises via a conditional expectation, from the corresponding posterior probabilities, in this case, of that stopping the trial due to futility would be a false conclusion.

These two ways of stopping the trial can be usefully complemented with a third possibility. It may turn out that neither efficacy nor futility has been concluded
even after the trial has been run for rather long, and then appear likely, based on an interim analysis of the data, that this situation would continue  even if many more patients would be enrolled, with the total costs thereby also accruing. This brings up the question of whether it would make sense to stop the trial even if the result would thereby remain \textit{inconclusive}. If so, what would be the logical point in time to make such a decision? 

This question of early stopping of the trial has been studied in a large body of the clinical trials literature, by employing a mix of traditional frequentist, Bayesian, and hybrid concepts, such as conditional power, predictive power  and probability of success. Contributions to the area include  \cite{Herson1979PredictivePE}, \cite{SPIEGELHALTER19868}, \cite{spiegelhalter1994bayesian}, \cite{geisser1994interim}, \cite{simon1994some} and \cite{johns1999use}, with the works \cite{dallow2011perils}, \cite{yi2012hybridization}, \cite{wiener2020methods}  and \cite{saville2023conditional} demonstrating a continued current interest in the topic. The advantages of the Bayesian approach to solving this problem have been explored, e.g., in  \cite{dmitrienko2006bayesian}, \cite{lee2008predictive}, \cite{saville2014utility} and \cite{https://doi.org/10.1002/sim.7685}, and more recently, in \cite{sambucini2021bayesian} and \cite{Beall2022InterpretingAB}. Particularly relevant to us are \cite{cheng2005bayesian}, where this question was studied for 2-arm trials from the perspective of Bayesian decision theory, and \cite{bassi2021bayesian}, where the approach was extended to multi-arm designs.

The structure of the paper is as follows. In Section \ref{section:no:2}, we present a general sequential rule for running a trial, and thereby for concluding  efficacy of the experimental treatment or futility of the trial. In Section \ref{subsection:no:2.3}, we show how the concerns of the regulators on false positives can be accounted for without using the concept of type I error rate, suggesting that it be replaced by an alternative criterion that respects the likelihood principle. In Section \ref{subsection:no:2.2} the decision rules of Section \ref{section:no:2} are extended to allow the trial to be stopped early, without having reached  a definitive result. Section \ref{section:no:4} contains some numerical illustrations. The paper  ends with a discussion in Section \ref{section:no:5}.

\section{Sequential rules for concluding efficacy or futility}
\label{section:no:2}

In a 2-arm \textit{superiority trial} the usual goal is to find out whether experimental treatment is  better than selected control, the latter representing a commonly used standard treatment. Here, for simplicity, incoming patients could be assumed to be assigned to the treatments according to a random block design of size two; the blocks are then independent and the order inside each block has been randomized, with equal probabilities, in advance. Also, more general sequential assignment rules such as response adaptive randomization (RAR), e.g., \cite{robertson2023response}, can be employed without changing the presentation below as long as such rules only depend on the past assignments and the corresponding outcomes, and not on the parameters of interest, here the true success rates of the two treatments. 

Considering, throughout, the prototype case of exchangeable dichotomous outcomes, we denote by ${\mathcal D}_{n}$ the data sequence consisting of treatment assignments and outcomes of the first $n$ patients, $n\ge1$. Exchangeability, a standard concept in Bayesian statistics and generalizing the more familiar i.i.d., expresses here an assumed symmetry property between the patients; the technical assumption is that, for any $n$, the distribution of ${\mathcal D}_{n}$ is not affected if the order in which these patients are treated is permuted in some way.

Instead of systematic monitoring of the data after every new outcome, such measurements are usually registered less often, at \textit{interim} times  after $\sigma_n$ patients have been treated, with $n \ge 1$ and $1 \le \sigma_1 < ... \sigma_{n-1} < \sigma_n.$ The sequence  $\{\sigma_n, n \ge 1\}$ would then be determined according to some fixed or inductive rule such that $\sigma_{n+1}$ is always determined
by the data ${\mathcal D}_{\sigma_n}$. 

In practice, if interim analyses are performed at all, their number is mostly small and timing fixed in advance in the design. Although potentially useful, interim analyses always require some degree of unblinding, complicate the logistics and increase the computational costs. However, as noted earlier, due to the reliance on the likelihood principle, interim analyses, if their timing is either fixed or determined inductively by the data, do not as such influence the statistical analysis of the data or the error analysis, as would be the case if the latter would be based on bounding the type I error rate.  In particular, concepts such as $\alpha$-spending functions are here irrelevant.

Denote by $ \bftheta _{0} $ and $\bftheta _{1} $ the model parameters, the unknown success rates of the dichotomous outcomes from, respectively, the control and the experimental arms. Given $( \bftheta _{0},\bftheta _{1}) $, the outcomes from the two arms are assumed to be independent, both i.i.d. given the respective parameter. Boldface notation is here an indication of that, in Bayesian modeling, they are treated as random variables. Let $\pi_e$ and $\pi_f$ denote the (joint) prior distributions for $(\bftheta _{0}, \bftheta _{1})$, approved, respectively, by the regulators and the trial sponsors. The subscripts $e$ and $f$ refer, as is explained below, to the primary role of the regulators in establishing  \textit{efficacy} of the experimental treatment, and that of the sponsors for concluding \textit{futility} of the trial. Accordingly, denote by  $\P_{\pi_{e}}$ and $\P_{\pi_{f}}$ the corresponding 
joint probabilities on the product space of the parameters $(\theta_0, \theta_1)$ and of possible data sequences $\{{\mathcal D}_{n};1 \le n \le N_{\max}\}$; here $N_{\max}$ is the finite maximal trial size, fixed as part of the design.

\textbf{Remark.} The idea of introducing and utilizing different prior distributions, representing different background information and attitudes of the interested parties, appears to have been first presented in \cite{kass1989investigating}, see also \cite{spiegelhalter1994bayesian} and  \cite{10.1214/088342304000000080}. Some aspects relating to the choice of the priors $\pi_{e}$ and $\pi_{f}$ are presented later in subsection \ref{subsection:no:3.2}.

Restating the earlier somewhat more carefully: The natural goal of a superiority trial is being able to demonstrate, based on trusted prior information and evidence provided by the trial data, that  the experimental treatment is with 'high' probability better than the control, possibly by a prespecified margin. Expressed more formally, we are led to considering the following sequential rule:
 
\textbf{Rule for concluding efficacy:} Compute  systematically  the posterior probabilities 
$ \P_{\pi_{e}}  ( \bftheta _{1} > \bftheta _{0}  + \Delta   \vert  {\bf D}_{\sigma_n}  )$
at the times  $\sigma_n$ an interim analysis is made, and stop the trial for concluded efficacy if 
$ \P_{\pi_{e}}  ( \bftheta _{1} > \bftheta _{0} +   \Delta   \vert  {\bf D}_{\sigma_n}  ) \ge 1 - \varepsilon_e$. 

In here, a positive value of the    \textit{minimal important difference (MID) $\Delta$} provides some extra protection to the control arm in such a comparison. 
If   no such extra protection is needed, one can simply choose $\Delta = 0.$  

 
The role of $\varepsilon_e$ is understood best by considering the complementary event $\{\bftheta _{1} \le \bftheta _{0} +  \Delta\}$, and thereby writing the above inequality in the equivalent form  $ \P_{\pi_{e}}  ( \bftheta _{1} \le \bftheta _{0} +   \Delta   \vert  {\bf D}_{\sigma_n}  ) < \varepsilon_e$. Thus, if it were decided to stop the trial based on the data $ {\bf D}_{\sigma_n}$, concluding efficacy would be a false positive with posterior probability less than $\varepsilon_e$. The primary interest of the regulators is that of guarding against \textit{false positives}, and is then expressed in the specification of an appropriate value for the threshold value $\varepsilon_e $.  
 This  threshold then represents  the  tolerance level for the probability of false positive conclusion, or of \textit{ false discovery}, which the regulators have deemed acceptable in the context.

A similar criterion can be specified for stopping the trial for \textit{futility}, to correspond to the interests of the sponsors: If a posterior probability $  \P_{\pi_{f}}  (  \bftheta _{1} - \bftheta _{0} \geq 0  \vert  {\bf D}_{\sigma_n})$ computed at the time of an interim analysis turns out to be below a given $\varepsilon_f$, say, the trial is stopped and   its futility is concluded.  The chances of the experimental treatment actually being better than the control are then deemed by the investigators to be  small enough to justify such a conclusion.

A smaller value of $\varepsilon_e$ reflects then a more conservative attitude towards concluding efficacy, and  a smaller value of $\varepsilon_f $ towards ending the trial due to declared futility. Although based on different concepts of probability and, indeed, on different statistical paradigms, $\varepsilon_e > 0$ and $\varepsilon_f > 0$  can be seen as having operationally similar roles as the bounds for the control of type I and type II error rates in frequentist hypothesis testing.

The regulators could generally be expected to be more conservative towards concluding efficacy than the investigators, so that 
\begin{align} \label{stoch:ordering:ineq}
\P_{\pi_{f}}  (   \bftheta _{1} - \bftheta _{0}  \leq \Delta   \vert  {\bf D}_{\sigma_n}   ) \leq \P_{\pi_{e}}  (   \bftheta _{1} - \bftheta _{0}  \leq \Delta   \vert  {\bf D}_{\sigma_n}  ).
\end{align}
A sufficient condition for this is that  the prior distributions  $\pi_e$ and $\pi_f$ for $\bftheta_1 - \bftheta_0$  satisfy the monotone likelihood ratio (MLR) property, e.g., \cite{Marshall2011}. We return to the issue of prior selection later in subsection \ref{subsection:no:3.2}, where also the simple argument leading to \eqref{stoch:ordering:ineq} is given. 

We assume that $\varepsilon_e +\varepsilon_f < 1$, which together with (\ref{stoch:ordering:ineq}) implies that the  defining conditions $ \P_{\pi_{e}}  (   \bftheta _{1} - \bftheta _{0}  \leq \Delta   \vert  {\bf D}_{\sigma_n}  ) < \varepsilon_e$ and  
$  \P_{\pi_{f}}  (   \bftheta _{1} - \bftheta _{0} \geq 0  \vert  {\bf D}_{\sigma_n}) < \varepsilon_f $ cannot  be simultaneously satisfied for the same ${\bf D}_{\sigma_n}$.   If either one of them holds for some $\sigma_n$, the trial is stopped and the other one will never materialize. In practice, $\varepsilon_e$ and $\varepsilon_f$ are selected to be small, and then such co-occurrence is ruled out even if  inequality (\ref{stoch:ordering:ineq}) may not hold.
   
To summarize, let
\begin{align}\label{eq:no:11} 
\bftau = \inf \{ 1 \leq \sigma_n \leq N_{\max}: \P_{\pi_{e}}  (    \bftheta _{1} - \bftheta _{0}  \leq \Delta   \vert  {\bf D}_{\sigma_n}) < \varepsilon_e  \text{ or }  \P_{\pi_{f}}  (\bftheta _{1} - \bftheta _{0} \geq 0  \vert  {\bf D}_{\sigma_n}) < \varepsilon_f\} \end{align}
be the time at which the trial is stopped for either of these two reasons. We then define the random variables 
\begin{align}\label{eq:no:1}
\bftau_e = \left\{
\begin{matrix}
\bftau 
&\text{ if } 
\P_{\pi_{e}}(\bftheta _{1} - \bftheta _{0}  \leq \Delta  \vert  {\bf D}_{\bftau}) < \varepsilon_e
\\  \infty & \text{ otherwise} \end{matrix}\right.  \end{align}
and 
\begin{align}\label{eq:no:2}
\bftau_f = \left\{
\begin{matrix}
\bftau 
&\text{ if }   
\P_{\pi_{f}}(\bftheta _{1} - \bftheta _{0}  \geq 0   \vert  {\bf D}_{\bftau}) < \varepsilon_f, \\ \infty & 
\text{ otherwise}\end{matrix}\right.\end{align}
Thus $\bftau, \bftau_e$ and $\bftau_f$ are \textit{stopping times} relative to   the observed treatment assignment and outcome histories $\{ {\bf D}_{\sigma_n} ;1 \leq \sigma_n \leq N_{\max}\}$. Clearly $\bftau = \bftau_e \wedge \bftau_f$. A finite value of $\bftau_e$ signals that the trial is stopped due to concluded efficacy, and a finite value of $\tau_f $ due to concluded futility.  
These two ways of stopping the trial can, by employing a decision function notation $\delta(.)$, be written as: 

\textbf{(D:i)} $\delta({\bf D}_{\sigma_n}) = d_e$: on observing a value $\bftau_e = \sigma_n \leq N_{\max}$, the trial is stopped and the experimental treatment is declared to be \textit{effective (superior)} relative to the control, allegedly because it is believed that $\{\theta _{1} - \theta _{0} > \Delta\}$ holds; then  $ \P_{\pi_{e}}  (  \bftheta _{1} - \bftheta _{0} > \Delta   \vert  {\bf D}_{\bftau_e} \cap \{\bftau_e < \infty\} ) \ge 1 - \varepsilon_e$. 

\textbf{(D:ii)} $\delta({\bf D}_{\sigma_n}) = d_f$: on observing a value $\bftau_f = \sigma_n \leq N_{\max}$, the trial is stopped and declared to have ended in \textit{futility}, allegedly because it is believed that $\{\theta _{1} - \theta _{0} < 0 \}$ holds; then  $ \P_{\pi_{f}}  (  \bftheta _{1} - \bftheta _{0} < 0   \vert  {\bf D}_{\tau_f}  \cap \{\bftau_f < \infty\}   ) \ge 1 - \varepsilon_f$. 

If neither $\bftau_e\leq \sigma_n$ nor $\bftau_f\leq \sigma_n$ holds, the trial is continued according to:

\textbf{(D:iii)} $\delta({\bf D}_{\sigma_n}) = d_c $: for $\sigma_n < \bftau = \bftau_e \wedge \bftau_f$, $d_c$ signifies the decision to continue the trial after $\sigma_n$  by enrolling and treating more patients until reaching the one indexed by $\sigma_{n+1}$. 

The stopping rule is then updated by replacing ${\bf D}_{\sigma_n}$ by ${\bf D}_{\sigma_{n+1}}$, i.e., by asking whether either $ \P_{\pi_{e}}  (    \bftheta _{1} - \bftheta _{0}  \leq \Delta   \vert  {\bf D}_{\sigma_{n+1}}  ) < \varepsilon_e$ or $\P_{\pi_{f}}  (  \bftheta _{1} - \bftheta _{0} \geq 0  \vert  {\bf D}_{\sigma_{n+1}}) < \varepsilon_f$ would hold. This inductive process can in principle be continued until reaching the maximal trial size   $N_{\max}$. If neither $d_e$ nor $d_f$ is concluded by then, the trial ends with:

\textbf{(D:iv)} $\delta({\mathcal D}_{N_{\max}}) = d_{\oslash}$: when neither $\bftau_e = \sigma_n $ nor $\bftau_f = \sigma_n $ is observed for some $\sigma_n \leq N_{\max}$, the trial ends at $\tau = N_{\max}$ by declaring its result to be \textit{inconclusive}.

Of course, these sequential decision rules can be employed only if the trial is not completely blinded. 

Once the trial has ended at $\bftau  \le N_{\max}$, the entire posterior distributions $ \P_{\pi_{e}}  (  (\bftheta _{0}, \bftheta _{1}) \in \cdot \space  \vert  {\bf D}_{\bftau} \cap \{\bftau < \infty\} ) $ and $ \P_{\pi_{f}}  (  (\bftheta _{0}, \bftheta _{1}) \in \cdot  \space\vert  {\bf D}_{\bftau} \cap \{\bftau < \infty\} ) $ can in principle revealed.
For example, one can compute, with respect to these posteriors, expected values of functions of suitably defined utility values. This is a major advantage compared to rules based on traditional hypothesis testing. Then, if the trial ends with a non-significant result, nothing beyond that simple fact is learned from the trial. 

An extension of \textbf{(D:iv)}, providing a utility based decision rule for an early stopping of the trial, is considered later in Section \ref{subsection:no:2.2}.

\section{Satisfying the requirements of the regulators }
\label{subsection:no:2.3} 
 
\subsection{Background }
\label{subsection:no:2.4}

Practical implementation of a trial design  always needs to be  approved by the responsible regulatory authorities. This creates a potential problem for Bayesian designs, because the standard requirement of the regulators is stated in frequentist terms, as an upper bound for type I error rate. 

For example, \cite{ventz2015bayesian} write as follows: "Concepts, such as the type I error probability, are required to be part of the study design to gain regulatory approval for a treatment", adding  later  "If necessary, the investigator adjusts tuning parameters to obtain a design that satisfies a pre-specified constraint on the type I error probability." This requirement of adjusting the parameters of a proposed Bayesian design,  to be compatible with a given bound for type I error rate,    is found also in guidelines such as \cite{fda_71512}, \cite{fda_78495} and \cite{ICH_E20}. \cite{zhou_ji_2023}, in their recent review paper on sequential Bayesian trial designs,  call this approach \textit{The Frequentist-oriented Perspective}. 

There are many examples in the statistical literature on how the settings of a Bayesian design might be adjusted to match the type I error bound, e.g., \cite{freedman1994and}, \cite{Grieve2016IdleTO}, \cite{nikolakopoulos2016hybrid}, \cite{doi:10.1177/0962280215595058},  \cite{shi2019control} and \cite{lee2024using}. However, strict enforcement of a numerical control rule, which has its origin in the frequentist tradition, not only necessitates the introduction of modeling apparatus that would not be necessary from a Bayesian perspective, but makes the Bayesian approach literally subordinate to frequentist control.
In particular, even though  the decision rule \textbf{(D:i)} itself respects the likelihood principle, the multiplicity problem and the consequent inflationary behavior of this criterion are then brought back, now as if through the back door. A more detailed discussion of this question is deferred to subsection  
\ref{subsection:no:3.2}. 

To correct this troubling situation in Bayesian trial designs, we make the bold suggestion of dropping type I error rate altogether from the methodological tool box, and replacing it by another criterion for error control. This criterion, introduced below, not only is consistent with Bayesian principles but actually follows directly, by a simple mathematical argument, from the decision rule \textbf{(D:i)} for concluding efficacy when analyzing the trial data. In other words, if the regulators accept the use of \textbf{(D:i)} as part of the trial design, the suggested criterion is satisfied automatically.

\subsection{Proposing a  criterion for the regulators' error control}
\label{subsection:no:3.3} 

The primary interest of the regulators, traditionally expressed in the form of bounding type I error rate, is that of controlling the occurrence of false positives. In the present case of a comparative 2-arm trial with dichotomous outcomes, type I error rate is interpreted as the relative proportion of erroneous claims of efficacy determined from a large number of trials, with the parameters  $\theta_1$ and $\theta_0$ fixed and such that  $\theta_1 = \theta_0$. One may wonder, however, whether this quantity is indeed the most logical criterion that should be controlled for such a purpose? Arguably, the real question of interest is not how often false positives under such circumstances may occur but, rather, that if a claim of efficacy is made, how likely is this to be false? In other words, how likely is it that, in actual fact,
$\{\bftheta _{1} - \bftheta _{0}  \le \Delta\}$ is true?


In view of this, and given that  trial designs are presented to the regulators for approval at a time at which no actual trial data are yet available, posterior probabilities of the form $ \P_{\pi_{e}}  (   \bftheta _{1} - \bftheta _{0}  \leq \Delta   \vert  {\bf D}_{\sigma_n}  ) $ employed in \textbf{(D:i)} cannot be determined.    Instead,  we are led to considering conditional  probabilities of the form 
\begin{align*}\P_{\pi_e}   (\bftheta _{1} - \bftheta _{0}  \le \Delta   \vert \bftau = \bftau_e < \infty) = \frac{ \P_{\pi_e}   (\bftheta _{1} - \bftheta _{0}  \le \Delta, \bftau = \bftau_e < \infty) }{ \P_{\pi_e} ( \bftau =  \bftau_e < \infty)}.  \end{align*} 
Here the conditioning event $\{\bftau = \bftau_e < \infty\}$ is that efficacy of the experimental treatment is, or will be, concluded in the trial, while $\{\bftheta _{1} - \bftheta _{0}  \le \Delta\}$ says that  its success rate  does not exceed that of the control by the required MID.  
This gives rise to the following

\textbf{Definition.} \textit{We call the conditional probability} 
\begin{align*}\P_{\pi_e}   (\bftheta _{1} - \bftheta_{0}  \le \Delta \vert \bftau = \bftau_e < \infty
)\end{align*} \textit{False Discovery Probability, abbreviating it as FDP}.

\textbf{Remarks.} (i) FDP and type I error rate are operationally similar in that a corresponding upper bound, $\varepsilon_e>0$ or  $\alpha>0$, is specified as part of the design. Their most obvious  difference, already alluded to above, is the direction of reasoning. For type I error rate, it is from postulating, in the parameter space, a condition of "no difference between treatments" and then  considering the possibility, in the sample space, of ending up with making a "discovery". For FDP, in contrast and as in all of Bayesian statistics, the starting point is that a "discovery" is made, and the focus is on
finding  possible explanations for this in the parameter space. Other differences are in how the underlying probability concepts for these two measures are to be understood and, to some extent, also in the numerical computations that are needed. 

(ii) FDP is closely related to a number of analogous concepts well known from the statistical literature. 
Due to its conditional form, it resembles the quantity $1 - $PPV, where PPV is the familiar  characteristic \textit{Positive Predictive Value} in laboratory testing, commonly evaluated from empirical data as the proportion of true positives from all positive results. The corresponding theoretical concept, a conditional probability, is determined by Bayes' rule from the prevalence of the condition in the population and the sensitivity and the specificity of the test method. PPV has also been used as an instrument in more general contexts, as in the discussion of why so many results from empirical research are erroneous (\cite{ioannidis2005most}).  FDP is similarly related to the criterion called \textit{False
Positive Report Probability} (FPRP), introduced in \cite{wacholder2004assessing} in the context of cancer genetics. In FPRP, the conditioning event is  a 'positive report', meaning that a 'statistically significant' result has been obtained from a hypothesis test between  alternatives $H_0$ and $H_1$ when using  significance level  $\alpha$ and power $1 - \beta$. The numerical value of FPRP is then determined by Bayes' rule, when also accounting for the prior probabilities of $H_0$ or $H_1$ being true. Some comments on the related concepts of \textit{False Discovery Rate} (FDR) (\cite{benjamini_1995}) and \textit{positive False Discovery Rate} (pFDR) (\cite{storey2003positive}) are provided later in \ref{subsection:no:3.2}. 

(iii) The conditional character in the above criteria 
is important also quantitatively: For example,  suppose that  one, in a series of trials, uses systematically tests based on $\alpha = 0.05$ and  $1 - \beta = 0.80$. Suppose further that only 1 in every 5 of investigated treatments is truly effective, corresponding to prevalence 0.2. A simple computation applying  Bayes' rule shows then that 20 percent of positive reports will be false,  fourfold  the selected $\alpha$-level. If only 1 in every 10 are truly effective, then 36 percent of the positives are false. For example, in drug development such instances may not be very rare, and then simple reliance on a selected small $\alpha$ will be deceptive. 

(iv) The event $\{\bftau = \bftau_e < \infty\}$ is of course uncertain at the time the trial design is being contemplated, and may not actually happen when the trial is run. But if no such 'discovery' is made, then 'false discovery' is not only unlikely but impossible. In this case, the conditioning event in defining FDP is not realized in the trial and its use remains mute.

(v) If a finite value of $\bftau_e$ is established,  the corresponding complete data sequence ${\bf{D}}_{\bftau_{e}}$ has been recorded, thereby giving rise to the   posterior probability  $\P_{\pi_e}   (\bftheta _{1} - \bftheta _{0}  \le \Delta   \vert {\bf{D}}_{\bftau_{e}} \cap \{\bftau = \bftau_e < \infty \} ).$ This probability is dominated by $\varepsilon_e$, because this is how $\tau_e$ was defined. Following well known rules of probability calculus, the FDP $\P_{\pi_e}   (\bftheta _{1} - \bftheta _{0}  \le \Delta   \vert \bftau = \bftau_e < \infty) $ can now be expressed as the expectation, with respect to $\P_{\pi_e} $ and given $\{\bftau = \bftau_e < \infty \}$, of such "more refined" random posterior probabilities. 


This observation now leads to the following key result:


\textbf{Proposition.} \textit{Suppose that the  regulators have approved a sequential trial design involving decision rule \textbf{(D:i)}, in which the trial is stopped for efficacy   at time $\bftau_e$ due to observing  
\begin{align*}\P_{\pi_e}   (\bftheta _{1} - \bftheta_{0}   \le \Delta \vert  {\bf{D}}_{\bftau_{e}} \cap \{\bftau = \bftau_e < \infty \})< \varepsilon_e.
\end{align*}Then also   \begin{align*}\P_{\pi_e}   (\bftheta _{1} - \bftheta _{0}  \le \Delta   \vert \bftau = \bftau_e < \infty) < \varepsilon_e .\end{align*}} 

\textbf{Proof.}  We get, by directly applying the arguments in (iv) above, that
\begin{align*} &
\P_{\pi_e}   (\bftheta _{1} - \bftheta _{0} \le \Delta    \vert \bftau   =\bftau_e < \infty) = \E_{\pi_e} (\P_{\pi_e}   (\bftheta _{1} - \bftheta_{0}   \le \Delta \vert  {\bf{D}}_{\bftau_{e}} \cap \{\bftau = \bftau_e < \infty \})   \vert \bftau = \bftau_e < \infty) & \\ &
<    \E_{\pi_e} ( \varepsilon_e \vert \bftau = \bftau_e < \infty) = \varepsilon_e.  \qed
\end{align*}

\textbf{Remarks.} 
(i) An important consequence of this result is that there is no need to perform the rather tedious numerical double integration  with respect to parameter values $(\theta_0,\theta_
1)$ and to data sequences $ {\mathcal{D}}_{\tau_{e}}$,  to determine the value of  the FDP $\P_{\pi_e}   (\bftheta _{1} - \bftheta_{0}  \le \Delta \vert \bftau =  \bftau_e < \infty )$, and then to find whether it would be below a selected error tolerance $\varepsilon_e $. This will be automatic if  decision rule \textbf{(D:i)} is applied systematically in all interim analyses, regardless of the particular data sequence  $\{{\mathcal{D}}_{\sigma_n}\}$ that may come up. 

(ii) The values for design parameters $\pi_e, \pi_f, \varepsilon_e, \varepsilon_f$ and $\Delta$ can be selected in a meaningful way only in a considered substantive context, in consultation between the sponsors and the regulators at the time the trial is being designed, thereby corresponding directly on their roles in the stopping rules \textbf{(D:i)} and  \textbf{(D:ii)}. In principle, once their values   have been selected in the   desired way and approved by the parties concerned, the trial can be started without, as is presently the common practice,    experimenting with preliminary simulations to determine an appropriate trial size. (In addition to stopping the trial due to concluded \textit{efficacy} or \textit{futility}, the possibility of  \textit{inconclusive} early stopping is considered in Section \ref{subsection:no:2.2}).

(iii) The decisions during the analysis are always based on the current posterior distribution, not on criteria such as FDP.  When more data become available, the conclusions  become less sensitive to the choice of the prior. After having actually  established that the trial has ended at a finite value of $\bftau_{e}$, and having thereby observed the   data $ {\bf{D}}_{\tau_{e}}$, the bound FDP $< \varepsilon_e$ for the probability of false discovery, guaranteed by the above Proposition, can be replaced by the realized value of $\P_{\pi_e}   (\bftheta _{1} - \bftheta _{0}  \le \Delta   \vert {\bf{D}}_{\tau_{e}}).$

(iv) The smaller the value of $\varepsilon_e$, the more outcome data are required before the conditioning event $\{\bftau = \bftau_e < \infty\}$ can be established, if at all, and the weaker is the dependence of     $\P_{\pi_e}   (\bftheta _{1} - \bftheta_{0}  \le \Delta \vert  {\bf{D}}_{\bftau_{e}} \cap \{\bftau = \bftau_e < \infty \})$  on the prior $\pi_e$. If the sensitivity aspect is a real concern to either party, the design can include a compulsory burn-in period such that the decision rules \textbf{(D:i)} - \textbf{(D:iv)} are activated only after the burn-in period has passed. Technically, this means that $\sigma_1$, the number of treated patients before initiating the first interim analysis, is selected to be sufficiently large. For a more general discussion of the impact of prior selection in Bayesian trial designs see, e.g., \cite{morita2010evaluating}.

(v) In the above analysis, a reported conclusion $d_e$ on efficacy  has been interpreted as being false if, in fact, $\{\theta _{1} - \theta _{0} \le \Delta\}$ is true, i.e., the experimental new treatment may be better than the control, but not by the required MID margin $\Delta > 0$. Another possibility, quite plausible in practice, would be to instead use the stronger criterion  $\{\theta _{1} - \theta _{0} \le 0\}$ for interpreting a declared efficacy conclusion to be false and then reserve the term FDP for $\P_{\pi_e}   (\bftheta _{1} - \bftheta _{0} \le 0 \vert \bftau = \bftau_e < \infty)$.   Since  \begin{align*}\P_{\pi_e}   (\bftheta _{1} - \bftheta _{0} \le 0 \vert \bftau = \bftau_e < \infty)  \le \P_{\pi_e}   (\bftheta _{1} - \bftheta _{0} \le \Delta    \vert \tau = \tau_e < \infty), \end{align*} the statement of the Proposition remains valid in this case as well. Yet another possibility  is to let simply $\Delta = 0$.  

(vi) Unsubstantiated conclusions regarding \textit{futility} are likely to be of much less concern to the regulators than false claims of efficacy, but they should be of interest to the investigators of the trial and the sponsors and stakeholders behind the development of the considered experimental  treatment. A declared futility result $d_f$ is unsubstantiated if in fact $\{\theta _{1} - \theta _{0} \ge 0\}$ is true, i.e., the considered new treatment is at least as good as the control. As an alternative, one could even require that the stronger condition $\{\theta _{1} - \theta _{0} \ge \Delta\}$, involving a MID $\Delta > 0$, be satisfied in order to make such a judgment.  Depending on the choice, one can consider conditional probabilities $\P_{\pi_f}   (\bftheta _{1} - \bftheta _{0} \ge 0 \vert \bftau = \bftau_f < \infty)$  and $\  \P_{\pi_f}   (\bftheta _{1} - \bftheta _{0} \ge \Delta    \vert \bftau = \bftau_f < \infty)$, calling the preferred one FFP for \textit{False Futility Probability}. In full analogy with the above Proposition, we then see that if a risk tolerance $\varepsilon_f$ is applied  in the analysis of the trial data, consistent with rule \textbf{(D:ii)}, we have  $\  \P_{\pi_f}   (\bftheta _{1} - \bftheta _{0} \ge \Delta    \vert \bftau = \bftau_f < \infty) \le \P_{\pi_f}   (\bftheta _{1} - \bftheta _{0} \ge 0 \vert \bftau = \bftau_f < \infty) \le \varepsilon_f$. In other words, the FFP value is automatically bounded by the selected  $\varepsilon_f.$

\subsection{Selection of priors in a trial design: a critical look }
\label{subsection:no:3.2} 

In their recent review on Bayesian sequential  trial designs,  \cite{zhou_ji_2023} divided the approaches in the literature into three classes, calling them \textit{The Subjective Bayesian Perspective},  \textit{The Frequentist-oriented Perspective} and 
\textit{The Calibrated Bayesian  Perspective}. Below, we follow this taxonomy, providing a short description of each, with some comments. 

However, we drop   the attribute \textit{subjective} from the first one.
Even if not intended, it is often understood as having a negative connotation, to mean 'not scientific' (cf.  \cite{O’Hagan29032019}) or even arbitrary, thereby indirectly giving the impression that statistical methods relying on frequentist ideas would be in some sense \textit{objective}, cf. \cite{efron1986isn}. Another reason is that, 
to be approved for practical implementation, a trial 
design, including its  prior, always needs to be assessed by several domain experts; it can therefore perhaps be described as \textit{inter-subjective}  but is never a single subject's guesstimate. Finally, one could question whether the  \textit{Frequentist-oriented Perspective} should be called Bayesian at all, as it involves elements that are in direct conflict with the foundations of Bayesian statistics.   

\textbf{The  Bayesian Perspective}. 
As such, the models in this category represent the current standard in most Bayesian statistical literature, where probability is understood as an epistemic concept without explicit
reference to frequency interpretation.   
  
The  choice of the prior is often considered to be  a controversial issue, as it can be seen either as an advantage, by providing an opportunity to bring existing subject matter knowledge into the inferential problem, or an inherent weakness of the Bayesian approach, since the results of the data analysis will then depend on how the prior is chosen. In consequence of the latter, many alternative approaches have been introduced, to establish ‘non-informative’, 'weakly informative' or even ‘objective’ prior distributions. 

Literature surveys on the elicitation of prior distributions can be found in \cite{buck2006uncertain} and  \cite{azzolina2021prior}. The latter  identified altogether 460 articles in this area  until Nov. 2020, of which 42  were concerned specifically with applications to clinical trials.  About 80 percent of them reported on employing parametric techniques for the elicitation. As prior elicitation is not our focus and the literature on the subject is vast, we give only some directions for further reading. A key contribution, introducing the idea of a "community" of priors, is \cite{spiegelhalter1994bayesian}, while \cite{doi:10.1177/1740774509356002} present some ideas on employing historical information on controls. \cite{dallow2018better} give useful practical advice for the construction of prior distributions, and the recent survey paper by \cite{mikkola2024prior} puts the task of prior elicitation into a wider perspective.

Our suggestion above has been to employ two distinct prior distributions in the design: $\pi_e$ to be reviewed and approved by the regulators for the purpose of
concluding \textit{efficacy},
and $\pi_f$ selected by the trial sponsors and investigators for  concluding \textit{futility}. Appropriate specification of such priors will naturally depend strongly on the context considered. Following \cite{10.1214/088342304000000080}, also \cite{kass1989investigating}, $\pi_e$ may perhaps be chosen to be “sceptical” in that it would "express scepticism about large treatment effects ...  and [be seen] as a way of controlling early stopping of trials on the basis of fortuitously positive results." The prior $\pi_f$ might  be similarly described as optimistic, or even "enthusiastic", "with a low chance ... that the true treatment benefit is negative." In addition to employing  different priors, one may naturally consider varying threshold values $\varepsilon_e$ and $\varepsilon_f$, see, e.g., \cite{fayers1997tutorial}.

Thereby the roles of $\pi_e$ and $\pi_f$ are  separated clearly from each other. If, however,  the different parties involved would have a similar understanding on the likely values of the effect sizes, they  could express this in terms of a consensus prior  $\pi_e =\pi_f$.

The practical elicitation of  appropriate priors is facilitated considerably  by  postulating that $\bftheta_0$  and  $\bftheta_1$  are  independent. For  $\pi_e$ this means that $\pi_e(\theta_0,\theta_1) = \pi_e^{(0)}(\theta_0)\pi_e^{(1)}(\theta_1)$, where $\pi_e^{(0)}$ is the prior used for the control arm and $\pi_e^{(1)} $ that for the experimental arm.  Prior independence is commonly justified by its  convenience; it allows for separate elicitation of the arm specific univariate priors, the corresponding   posteriors can be updated based on only outcome data from that same arm, and finally, the independence of $\bftheta_0$  and  $\bftheta_1$ will still hold with respect to the consequent joint posterior. 

Prior independence of $\bftheta_0$  and  $\bftheta_1$ with respect to  $\pi_e$ would appear natural particularly in  confirmatory comparisons of the two treatments during phase III, on the grounds that the outcomes from one arm then do not influence how the performance of the other arm will be assessed. Moreover, the size of confirmatory trials is typically large, in which case the updated data likelihood, being a product of two Binomial likelihoods, will soon dominate the influence of the prior. 

The situation can be somewhat different, particularly in exploratory analysis during phase II. Outcome data from the experimental treatment are often sparse but there may be relevant historical data on the control. Then,  if the two treatments are by domain experts judged to be sufficiently similar, prior dependence across the arms may be a useful option to consider as it will allow for 'borrowing strength' across the two arms. There are many different ways of incorporating such dependence structures into the priors, for example,   by applying odds ratios or log odds (\cite{kass1989investigating}, \cite{spiegelhalter2004bayesian}),    hierarchical models  (\cite{BERRY1999215}),  mixtures computed from elicited histograms (\cite{thall2019bayesian}), or   partial exchangeability (\cite{diaconis2023approximate}). 

By far the most popular prior distribution  arising from dichotomous outcomes is Beta$ (\theta; a,b)$, where it is conjugate to the binomial likelihood. Moreover, for integer valued parameters the sum $a + b$ is considered as expressing the strength of the prior information, equivalent to $n=a + b$ observed outcomes of which $a$ are successes and $b$ are failures. If two Beta-priors, say Beta$ (\theta; a,b)$ and Beta$(\theta; a^*, b^*)$, satisfy $a \ge a^*$ and $b  \le b^*$, we can readily think of the former specifying a \textit{stochastically larger} random variable than the latter. In fact, the two priors satisfy then the stronger    \textit{likelihood ratio ordering} condition (see, e.g.,   \cite{Marshall2011}). Moreover, if 
the same data likelihood is used for both, likelihood ratio ordering is naturally inherited from priors to posteriors. This brings us back to inequality \eqref{stoch:ordering:ineq}  considered in Section \ref{section:no:2}.

Finally recall that, due to de Finetti's Representation Theorem, one of the corner stones of Bayesian statistics, the \textit{existence} of a prior distribution is guaranteed by merely postulating infinite exchangeability, under an assumed probability $\P$,  of the  outcomes from the considered trial. The likelihood component in this representation has the same canonical binomial form regardless of whether  Bayesian or frequentist ideas and methods are employed in the design (cf. \cite{arjas2016wearing}).

\textbf{The Frequentist-oriented Perspective}.  Here, as already mentioned in  \ref{subsection:no:2.4}, the idea is to apply a Bayesian model  in parameter estimation but then trust in traditional frequentist criteria in error control. 
To facilitate  comparison to the presentation in \cite{zhou_ji_2023}, consider the simple situation,   where $\bftheta_0$ is fixed at a given level $\bftheta_0 = \theta_0$, effectively reducing the  2-arm setting  to a single-arm trial. When expressed in the hypothesis testing framework, letting the MID $\Delta =0$ and dropping the redundant subscript from   the parameter $\theta_1$ of the experimental treatment, we are testing $H_0: \theta \le \theta_0$  against $H_1: \theta > \theta_0$.   In addition,  a  value $\theta_a > \theta_0$ is selected to  represent the composite alternative hypothesis $H_1$. Denote by $\P_{\theta_0}$and $\P_{\theta_a}$ the probabilities for which the outcome data are generated, respectively, by the fixed parameter values $\theta_0$ and $\theta_a$.

Suppose now, as we are concerned with sequential Bayesian designs, that the trial is run by following the rules \textbf{(D:i)} - \textbf{(D:iv)}, with $\pi_e$ as the prior   for $\bftheta$.
For the purposes of error control  in a design following the \textit{Frequentist-oriented Perspective}, type I error would be made if, in a trial in which outcome data would be consistent with $\P_{\theta_0}$, application of \textbf{(D:i)} - \textbf{(D:iv)} would lead to concluding efficacy, i.e., establishing a finite value of  $ \bftau_e$. To make the dependence of  $ \bftau_e$ on $\pi_e$ and  $\varepsilon_e$  explicit, also in situations in which data come from either $\P_{\theta_0}$ or $\P_{\theta_a}$, denote  it  by   $\bftau_e^{\pi_e,\varepsilon_e}$. 
  
A 'Bayesian' type I error rate can now be written as the probability $\P_{\theta_0}(\bftau_e^{\pi_e,\varepsilon_e}
< \infty )$. Thus, according to the \textit{Frequentist-oriented Perspective}
it is required that, for given significance level $\alpha>0$,  $\P_{\theta_0}(\bftau_e^{\pi_e,\varepsilon_e} < \infty ) \le \alpha$. This condition is commonly complemented with $\P_{\theta_a}(\bftau_e^{\pi_e,\varepsilon_e} < \infty ) \ge 1 - \beta$ for power. 

With $\alpha$ and $\beta$ fixed, these inequalities are not necessarily satisfied by the prior $\pi_e$ and the threshold $\varepsilon_e$ that would be selected when following the \textit{Bayesian Perspective}. In this case, they are to be adjusted in some way, as demanded by these frequentist criteria. As noted earlier, this makes the Bayesian analysis of trial data literally subordinate to the selected criteria of the frequentist error analysis. There is an extensive literature, of which some examples were provided earlier in  \ref{subsection:no:2.4}, on how such adjustments can be achieved technically. 

Such forcing calls for comments. From a Bayesian perspective, adjustments of the threshold values $\varepsilon_e$ would perhaps appear to be more acceptable than those  modifying the prior $\pi_e$,
as the former would not change the probability $\P_{\pi_e}$ when applying rule \textbf{(D:i)},  possibly only tightening the requirement for concluding efficacy. In contrast, changing the prior, only  to satisfy a numerical condition stemming from a different statistical paradigm, would conflict with the core ideas of Bayesian inference and decision theory.

However, requiring that the $\varepsilon_e$-values be adjusted due to bounding type I error rate by a selected $\alpha>0$ is not as innocent as it might seem. The reason is that, in sequential decision making, the well-known phenomenon of multiplicity sets again in, with the effect of either inflating  type I error rate or lowering the power. Thereby one of the key advantages from following the likelihood principle, viz. that the same threshold value can be used in error control regardless of how many interim analyses are made, is lost.

Note that the  computation of the numerical values of  $\P_{\theta_0}(\bftau_e^{\pi_e,\varepsilon_e} < \infty )$ and $\P_{\theta_a}(\bftau_e^{\pi_e,\varepsilon_e} < \infty ) $, which are needed in setting up a \textit{frequentist-oriented}  Bayesian design, to check that the former is below but close to $\alpha$ and the latter at least $1 - \beta$, requires  tedious double integrations, inside   with respect to  parameter values $\theta$ in order to evaluate posterior probabilities, and outside with respect to data sequences ${\mathcal{D}}_{\sigma_n}$ sampled from $\P_{\theta_0}$ and $\P_{\theta_a}$. Only integrations of the former type are needed when following the \textit{Bayesian Perspective}, and they are then conditioned on the trial data ${\bf{D}}_{\sigma_n}$ observed at interim times $\sigma_n$ when the trial is run.


\textbf{The Calibrated Bayesian Perspective}.
In their review paper, \cite{zhou_ji_2023} provided the following justification to this approach: "Although Bayesian probabilities represent degrees of belief in some formal sense, for practitioners and regulatory agencies, it can be pertinent to examine the operating characteristics of Bayesian designs in repeated practices." By operating characteristics, these authors refer to "the long-run average behaviors of a statistical procedure in a series of (possibly different) trials." The crux here is in the words in  parentheses, because the usual operating characteristics, type I and type II error rates, are interpreted as arising from an imaginary series of repetitions of the \textit{same} trial, with fully specified fixed parameter values. 

According to the \textit{Calibrated Bayesian Perspective} outlined in \cite{zhou_ji_2023}, each trial proposed to the regulators for approval is viewed as a random draw from a population of comparable trials "in repeated practices" referred to above, with the variation of treatment success parameters  $\bftheta_i$ in consecutive trials then characterized   by a  population distribution denoted by $ \pi_0.$ In one way or another, it is thought that $\pi_0$ would have an objective existence, and the Bayesian prior $\pi_e$ used in the analysis of the trial data should then be \textit{calibrated} accordingly. 

For this purpose, considering the example of a single-arm trial,  \cite{zhou_ji_2023} introduced two operating characteristics, which they called  \textit{False Discovery Rate} (FDR) and \textit{False Positive Rate} (FPR). In the words of \cite{zhou_ji_2023}, "The FDR is the relative frequency of false rejections among all trials in which $H_0$ is rejected, and the FPR is the relative frequency of false rejections among all trials with nonpositive treatment effects $\theta$s." 

Extensive simulation experiments were carried out in \cite{zhou_ji_2023} (in the case of conjugate Normal-Normal models with common mean) by varying both $\pi_0$ and $\pi_e$. The results, using FDR, FPR and coverage as performance criteria, were best when $\pi_e =\pi_0$. In this case the selected threshold value corresponding to our $\varepsilon_e$ was upheld overall, in agreement with our Proposition in \ref{subsection:no:3.3}. The more $\pi_e$ and $\pi_0$ differed from each other, the more the results deteriorated, and more so if the number of interim analyses was increased. 

 One may well ask, in what sense the population distribution $\pi_0$ could be said, excepting simulation experiments, to be 'known'? But if it were, why then not, justified by the idea that success parameter $\bftheta_i$ of a currently considered trial can be  thought of as "a random draw from the population distribution $\pi_0$", make the obvious choice  $\pi_e =\pi_0$ for the prior? If this is done, the consequent mathematics coincides with that of the \textit{Bayesian Perspective.} In particular,   FDR  coincides mathematically with FDP, in this single-arm case, the conditional probability $\P_{\pi_e}(\bftheta \le \theta_0 \vert \bftau = \bftau_e < \infty).$  
 Similarly, FPR  can be identified with \textit{False Positive Probability} (FPP), i.e., the conditional prior predictive probability  $\P_{\pi_e}(\bftau = \bftau_e < \infty \vert  \bftheta \le \theta_0 )$ of falsely concluding efficacy of the experimental treatment, given that its success rate does in fact not exceed the target $\theta_0.$ The third letters 'R' and 'P' in the acronyms FDR and FPR vs. FDP and FPP correspond to their respective frequentist 'rate' and Bayesian 'probability' interpretations. 
 
\textbf{Remarks.} (i) The term \textit{False Discovery Rate} (FDR), adopted here from \cite{zhou_ji_2023}, corresponds to the concept \textit{positive False Discovery Rate} (pDFR) introduced in \cite{storey2003positive}, rather than to the widely used FDR characteristic of \cite{benjamini_1995}.  Interestingly, as shown in \cite{storey2003positive}, pFDR  can in the case of a sequence of i.i.d. hypothesis tests be equivalently expressed as a conditional  probability. As \cite{storey2003positive} remarked, "...  pFDR is quite flexible in its interpretation. This is especially appealing in that it ... can be used by both Bayesians and frequentists." Such flexibility in the usage of the dual concepts of probability and rate seems useful as it offers a bridge for communication in situations where the sponsors may prefer applying specifically Bayesian ideas and methods in their trial design and analysis, but where the regulators are inclined to follow more traditional frequentist criteria in their assessment. 

(ii) The term \textit{Calibrated Bayesian}, introduced by R. J. Little, is used also in a wider meaning. According to \cite{doi:10.1198/000313006X117837}, "inferences under a particular model should be Bayesian, but model assessment can and should involve frequentist ideas". In the present case of dichotomous outcomes,  the likelihood obtains the canonical binomial form 
so that the aspect of "model assessment" can realistically only concern the elicitation of the prior. As already alluded to above in our discussion of the \textit{Bayesian Perspective}, there are many different approaches, most of them employing parametric techniques, that can be applied usefully and in relatively simple ways for such a purpose. In addition, a version of \textit{Calibrated Bayesian} activity assumes a natural role during phase 4, when  a drug's safety and effectiveness are assessed after it has been already  approved and is on the market, by comparing data from long-term studies, and possibly from different patient populations, to the predictions based on the results from phase 3. Standard frequentist methods can then be used routinely for such a purpose.

\section{A rule for stopping the trial early with an inconclusive result}
\label{subsection:no:2.2}

The sequential scheme presented in Section \ref{section:no:2} offers much flexibility in running a clinical trial, and especially compared to traditional designs using a fixed sample size. However, it may well be that neither efficacy nor futility  can be established for even relatively large values of $N_{\max}$. This is quite likely to be the case if the true success rate  of   the   experimental arm is slightly higher than that of the control, but not by the required MID margin $\Delta$. 

This calls for an option, in which there would be a possibility for stopping the trial "early" at times $\sigma_n < N_{\max} $ when neither  $d_e$ nor $d_f$ have so far been concluded and the chances of reaching either one before $N_{\max}$ appear thin.  For this, like many other works earlier (e.g., \cite{johns1999use}, \cite{dmitrienko2006bayesian}, \cite{lee2008predictive}, \cite{saville2014utility} and \cite{https://doi.org/10.1002/sim.7685}), we apply ideas from decision theory and predictive inference. Our criterion for early stopping is similar to that employed in \cite{cheng2005bayesian}, but differs from it markedly in error control, which in \cite{cheng2005bayesian} is based on frequentist principles.  

The intuition behind such a predictive reasoning is obvious: If, at a considered time $\sigma_n$, and if the trial would be continued, the predicted \textit{gains} would exceed the predicted \textit{losses}, then there is a rationale for continuing the trial past $\sigma_n$, and otherwise not. 

\textbf{Specification of utility values.} 
Small values of $\varepsilon_e$ and $\varepsilon_f$ guard against making the conclusions $d_e$ and $d_f$ on false grounds. However, the ultimate purpose of the trial is not to reject but to perform well reasoned selection: conclude $d_e$ when the experimental treatment is truly better than the control, and $d_f$ when it is not better. In consequence, one should consider both the benefits and the costs that will arise from a trial. A good design is one in which the expected benefits are larger than the expected costs. Technically this leads to considering the trial design as a decision problem, with an appropriately defined utility function.

Let $G_e$ be the \textit{gain} from concluding $d_e$  when it is correct, and  $L_e$  the \textit{loss} (as an absolute value) when it is false. Similarly, let $G_f$ be the \textit{gain} from concluding $d_f$ when it is correct, and $L_f$ the \textit{loss} (as an absolute value) when it is false.  In addition, we introduce a \textit{cost} that  incurs from considering (recruiting and treating) each additional patient in the trial.

The choice of appropriate numerical values for $G_e$, $G_f$, $L_e$ and $L_f$ will depend on the concrete context considered, and would perhaps most naturally correspond to the financial interests of the stakeholders of the pharmaceutical company which has been responsible for developing the new experimental treatment. Realistically, all these values should be positive, making $-L_e$ and $-L_f$ negative. This holds even for the gain $G_f$ from correctly concluding futility, however disappointing such a conclusion may be from the perspective of the original goals set for the trial. The point is that, if the new treatment does not have the desired target performance, it is better to find this out. Very likely, to justify designing and running a trial at all, $G_e$ should be quite large, and possibly larger than any of these other values. $L_f$ can be quite large as well if the development costs of the new treatment have been high. In confirmatory phase 3 trials the investments, as well as the expected profits from success, are likely to be much larger than in phase 2, and this then reflected in how the utility values are chosen.   

Expressing such consequences of the decisions $  d \in \{d_e, d_f, d_c, d_{\oslash}\}$ in terms of corresponding \textit{utility functions} $U_n(d, (\theta_0,\theta_1))$, $0 \leq n \leq N_{\max}$,  then gives,  consistent with \textbf{D(i) - D(iv)}, the following:
 
\textbf{(U:i)} $  U_{n}(d_e, (\theta_0,\theta_1)) = G_e 1_{ \{\theta_{1} - \theta _{0} > \Delta   \}} -  L_e 1_{\{\theta_{1} - \theta _{0} \leq \Delta   \}}$;  

\textbf{(U:ii)} $ U_{n}(d_f, (\theta_0,\theta_1)) = G_f 1_{\{\theta _{1} - \theta_{0} < 0\}} -  L_f 1_{\{\theta _{1} - \theta_{0} \geq 0 \}}$; 

\textbf{(U:iii)} $  U_{n}(d_c, (\theta_0,\theta_1)) = - 1 $ for all $(\theta_0,\theta_1);$

\textbf{(U:iv)} $  U_{n}(d_{\oslash}, (\theta_0,\theta_1)) = 0 $ for all $(\theta_0,\theta_1).$

In here, $- 1$ is the negative \textit{cost} that incurs from enrolling and treating an additional patient. If the costs of the two treatments considered are markedly different,  \textbf{(U:iii)} can be modified accordingly. The gains $G_e$ and $G_f$, and the losses $ L_e$ and $L_f$, are supposed to be measured in the scale of this unit cost. 

Another possible refinement concerns \textbf{(U:iv)}: the \textit{inconclusive} result $d_{\oslash}$  may have considerable practical value that would justify assigning a positive utility value to it. For example, it can be useful, from the perspective of comparable experiments planned in the future, to output summaries from the posterior distribution of $(\bftheta_0,\bftheta_1)$ given the final data. This differs from designs based on standard significance testing. Then, as already mentioned earlier, if the trial ends with a non-significant result, the statistical machinery is silent on anything but the fact that the trial was unsuccessful. A $p$-value larger than the selected $\alpha$ cannot be viewed as providing support to a Null hypothesis. 

Since the true parameter values $(\theta_0,\theta_1)$ are unknown, in a decision problem the corresponding utility function values need to be replaced by their posterior expectations at the considered time point. 

For this, suppose that neither $\bftau_e$ nor $\bftau_f$ have been observed by the time data ${\bf D}_{\sigma_{n-1}}$ from the $\sigma_{n-1} $ patients have been registered. In this case, only decisions $d_c$ have been made so far and, according to specifications above, the cumulative utilities have decreased with slope $- 1$ to value $-\sigma_{n-1}$. Consider then the situation at time $\sigma_n$ after having observed the data  ${\bf D}_{\sigma_n}.$
According to the above scheme: 
\begin{align*} 
\{\bftau_e = \sigma_n \} = \{\bftau \geq \sigma_n, \delta({\bf D}_{\sigma_n}) = d_e\}  =   \{\bftau \geq \sigma_n, \P_{\pi_{e}}  (    \bftheta_{1} - \bftheta_{0}  \leq \Delta   \vert  {\bf D}_{\sigma_n}) < \varepsilon_e\}.\end{align*}
The posterior expected utility from  concluding the experimental treatment to be effective relative to the control,  given $ {\bf
D}_{\tau_e}$, is therefore  
\begin{align}\label{eq:no:3}
\E_{\pi_{f}}( U_{\bftau_e}(d_e, (  \bftheta_{0}, \bftheta_{1})) \vert  {\bf D}_{\bftau_e}  )  =  G_e  - (L_e + G_e) \P_{\pi_{f}}  ( \bftheta_{1} - \bftheta_{0}  \leq \Delta    \vert 
{\bf D}_{\bftau_e}).
 \end{align}
Note that this expectation is taken with respect to $\P_{\pi_{f}}$, since the utilities are meant to correspond to the gains and losses of the sponsors and possible other stakeholders in question, although $\bftau_e$ appearing in the conditioning on the right is based on considering $\P_{\pi_{e}}$-probabilities. The expectation is positive if, as is natural to assume, $L_e < G_e$ and $\P_{\pi_{f}}  ( \bftheta _{1} - \bftheta _{0}  \leq \Delta    \vert  {\bf D}_{\tau_e})$ is small. The latter follows from the defining condition \eqref{eq:no:1} if
\begin{align*}
\P_{\pi_{f}}  ( \bftheta _{1} - \bftheta _{0}  \leq \Delta    \vert  {\bf D}_{\sigma_n}) \le \P_{\pi_{e}}  ( \bftheta _{1} - \bftheta _{0}  \leq \Delta    \vert  {\bf D}_{\sigma_n}).
\end{align*}
For futility we get
\begin{align*} 
\{\bftau_f = \sigma_n \} = \{\bftau \geq \sigma_n, \delta({\bf D}_{\sigma_n}) = d_f\} =   \{\bftau \geq \sigma_n, \P_{\pi_{f}}  (  \bftheta_{1} - \bftheta_{0} \geq 0  \vert  {\bf D}_{\sigma_n}) < \varepsilon_f\},
\end{align*}
and 
\begin{align}\label{eq:no:4} 
\E_{\pi_{f}}(
U_{\bftau_f}( d_f 
, (  \bftheta_{0}, \bftheta_{1})) \vert  {\bf D}_{\bftau_f})   
= G_f  - (L_f + G_f) 
\P_{\pi_{f}}  
(   \bftheta_{1} - \bftheta_{0} \geq 0 \vert  {\bf D}_{\bftau_f}). 
\end{align}
This expectation is positive if $\varepsilon_f \leq G_f/(L_f + G_f)$. Here, it is likely that the selected loss value  $L_f$ would be larger than the corresponding gain $G_f$.

Note that \eqref{eq:no:3} and \eqref{eq:no:4} involve convex combinations of the gains and the losses, resulting from the respective decisions $d_e$ and $d_f$ being correct or not correct, both weighted with the corresponding posterior probabilities of the sponsors. 
 
The smaller the values of the threshold values $\varepsilon_e$ and   $\varepsilon_f$ are, the longer it takes on average until either $d_e$ or $d_f$ can be established  based on the accumulated data. This is then reflected in a corresponding increase of the cumulative treatment costs and, when $N_{\max}$ is fixed, in bigger chances of ending the trial with the inconclusive result   
$\delta({\mathcal D}_{N_{\max}}) = d_{\oslash}$.

Let $T$   be the   \textit{time horizon} up to which, from the present time $\sigma_n$, such predictions are considered, $\sigma_n < T \leq  N_{\max}$. A possible choice would be to fix the length of the interval over which the prediction is to be made at some appropriate value, say $m$, in which case $T = (\sigma_n + m)\wedge N_{\max}$, and another to let that interval extend all the way to $T = N_{\max}.$   For any time point $t$ such that $\sigma_n \leq t \leq T $, we obtain the posterior predictive probabilities
\begin{align}\label{eq:no:5}
\P_{\pi_{f}}   (\bftau_e = t \vert  {\bf{D}}_{\sigma_n}\cap \{\bftau > \sigma_n \}) =
 \P_{\pi_{f}}   (\bftau \geq t,  \P_{\pi_{e}}  (  \bftheta _{1} - \bftheta _{0}  \leq \Delta   \vert {\bf{D}}_{t}) < \varepsilon_e \vert  {\bf D}_{\sigma_n}\cap \{\bftau > \sigma_n \})
\end{align}  
and 
\begin{align}\label{eq:no:6}
\P_{\pi_{f}}   (\bftau_f = t \vert  {\bf D}_{\sigma_n}\cap \{\bftau > \sigma_n \})  =  \P_{\pi_{f}}   (\bftau \geq t,  \P_{\pi_{f}}  (  \bftheta _{1} - \bftheta _{0}  \geq 0   \vert {\bf{D}}_{t}) < \varepsilon_f \vert  {\bf D}_{\sigma_n}\cap \{\bftau > \sigma_n \}).
\end{align}
Note that ${\bf{D}}_t$, appearing as a conditioning variable inside the right-hand expressions, is a random variable as it extends into "future" times $t>\sigma_n$ and is only  partly determined by the observed data ${\bf D}_{\sigma_n}$. The corresponding predictive expected cumulative utility, from $\sigma_n$ onward to $T,$ is then obtained by adding these two expressions for all $t$, when also accounting for the costs that will arise if neither $\bftau_e=t$ nor $\bftau_f=t$ is established in ${\bf{D}}_{t}$. This gives, with some computation,
\begin{align}\label{eq:no:7} &
\E_{\pi_{f}} \biggl(\sum _{t=\sigma_n}^{ \bftau \wedge T}U_t(\delta({\bf{D}}_t), (  \bftheta _{0}, \bftheta _{1})) 
\bigg\vert  {\bf D}_{\sigma_n}\cap \{\bftau > \sigma_n \} \biggr) 
& \\   \nonumber
&= \E_{\pi_{f}}\biggl(\sum _{t=\sigma_n}^{ \bftau \wedge T}1_{\{\P_{\pi_{e}}  (     \bftheta _{1} - \bftheta _{0}  \leq \Delta \vert  {\bf{D}}_{t}) < \varepsilon_e\} } 
\bigl[ G_e  - (L_e + G_e) \P_{\pi_{e}}  ( \bftheta _{1} - \bftheta _{0}  \leq \Delta    \vert  {\bf{D}}_{t}) \bigr]\bigg\vert  {\bf D}_{\sigma_n}\cap \{\bftau > \sigma_n \}  \biggr) 
& \\ & \nonumber
+  \E_{\pi_{f}}\biggl(\sum _{t=\sigma_n}^{ \bftau \wedge T}1_{\{\P_{\pi_{f}}  (   \bftheta _{1} - \bftheta _{0} \geq 0  \vert  {\bf{D}}_{t}) < \varepsilon_f\} }
\bigl
[G_f  - (L_f + G_f) \P_{\pi_{f}}  (   \bftheta _{1} - \bftheta _{0} \geq 0 \vert  {\bf{D}}_{t})  \bigr]
\bigg\vert  {\bf D}_{\sigma_n} \cap \{\bftau > \sigma_n \}\biggr) 
&\\ &\nonumber
- \biggl(\E_{\pi_{f}} (\bftau \wedge T\vert  {\bf D}_{\sigma_n}\cap \{\bftau > \sigma_n \} ) - \sigma_n  \biggr).
&\end{align} 
Numerical values of this predictive expectation can be computed by performing forward simulations from the posterior predictive distribution, given the data $ {\bf D}_{\sigma_n}$, of future trial developments ${\bf{D}}_{t}$ up to $t = \bftau \wedge T$, and finally by applying Monte Carlo averaging to compute the expectation. Due to the double integration, with respect to both the model parameters and the trial developments, these computations are much slower than those needed for computing the posterior probabilities  in Section \ref{section:no:2}. Realistically, such analyses would be made in practice only rarely.

According to the rationale presented above, we now have the following rule for stopping the trial with an inconclusive result:

\textbf{Rule for inconclusive stopping.} \textit{For a selected $m\ge 1$, let the running time horizon be $T_n= (\sigma_n+m)\wedge N_{\max}$, and
\begin{align} \label{eq:no:12}
\bftau_{\oslash} = \inf \{ &0 \leq \sigma_n \leq N_{\max}: \tau_e\wedge\tau_f\ge\sigma_n
& \\  \nonumber
&\text{ and } \E_{\pi_{f}} (U_{\bftau_e \wedge \bftau_f \wedge T_n}(\delta({\bf{D}}_{\bftau_e\wedge \bftau_f\wedge T_n}), (  \bftheta _{0}, \bftheta _{1}))
-(\bftau_e \wedge
\bftau_f\wedge T_n-\sigma_n)\vert  {\bf D}_{\sigma_n})  <0 \}. 
&\end{align}
If a finite value  $\bftau_{\oslash}=\sigma_n \leq N_{\max}$ is observed, the trial is stopped, thereby deciding  $\delta({\bf D}_{\sigma_n}) = d_{\oslash}.$ } 

This rule then extends and replaces the earlier definition {\bf{(D:iv)}} of $d_{\oslash}$. We also revise the meaning of our earlier notation $\bftau= \bftau_e \wedge \bftau_f$ into $\bftau= \bftau_e \wedge \bftau_f\wedge \bftau_{\oslash}.$

\textbf{Remarks.} (i) Note that essentially the same procedure, which is used for computing the predictive expectation \eqref{eq:no:7}, can be applied for computing quantities such as predictive probability of declared efficacy, or of futility, which are considered in a large body of literature, e.g., \cite{dmitrienko2006bayesian}, \cite{rufibach2016bayesian}, \cite{saville2014utility}, and many references therein. For computing the former, it suffices to insert  $ U_{n}(d_e, (\theta_0,\theta_1)) = 1_{\{\theta_1\ - \theta_0 \ge \Delta  \}}$ and $ U_{n}(d_f, (\theta_0,\theta_1)) =  U_{n}(d_c, (\theta_0,\theta_1)) =  U_{n}(d_{\oslash}, (\theta_0,\theta_1)) = 0  $ into expression \eqref{eq:no:7}, and finally ignore its last term, representing the treatment costs.

(ii) With some effort spent on computation, one can explore how different choices of the horizon $T$ would influence the numerical values of the expectation \eqref{eq:no:7}, and then perhaps choose one that would give it the maximal value. Note also that, in this scheme, no utility value can ever be larger than $G_e$, the gain from correctly declared efficacy
of the new experimental treatment. Since the costs accumulate at the rate of one unit per treated patient, the time horizon after $\sigma_n$ during which the utility remains positive is bounded by $G_e$ even when no finite $N_{\max}$ has been assumed; for a formal argument, see Theorem 1 in \cite{cheng2005bayesian}.
  
(iii) In the above scheme it is assumed that, in the simulated future trial histories ${\bf{D}}_t$ beyond the present time  $\sigma_n$, no further comparisons are made between the options $d_c$ and $d_{\oslash}$. This simplifies the argument and the consequent numerical computations considerably, as otherwise one would be led to considering a sequence of nested decision problems 
and thereby to applying recursive backwards induction to evaluate the expression numerically (e.g., \cite{carlin1998approaches}, \cite{wathen2006implementation}). On the other hand, several "present" times $\sigma_n$ for interim analyses can be fixed in advance, or specified as stopping times, as part of the trial design.  

(iv) The above rule for inconclusive stopping is in principle valid also at $n=0$, when  contemplating  whether starting a trial for testing   a new experimental treatment would be worth the effort and the costs that would thereby incur. At that time no outcome data are yet available from the study itself, and a careful elicitation of the prior is needed to arrive at a meaningful decision concerning such initiation.  

\section{Numerical illustrations of the characteristic features of the design}
\label{section:no:4}

In the statistical literature on Bayesian trial designs,  the performance of a proposed new method is commonly  demonstrated by simulation based numerical results. Of particular interest are then the standard operating characteristics: type I error rate, computed by Monte Carlo simulations where the model parameter values have been selected according to a Null hypothesis of "no difference between the treatments", and power or type II error rate, where one or more selected values for such differences are considered. Extensive illustrations of this kind can be found, for example, in the Supplement of \cite{Arjas2022}. 

Instead of simulations based on assumed known values of the treatment effect parameters, we consider below  the following three more general parameter configurations: (a) $\{\theta_1 - \theta_0 > \Delta\}$: new treatment better, by at least  the MID $\Delta$, than the control; (b) $\{\theta_1 - \theta_0 < 0\}$: new treatment worse than the control; (c) $\{0 \le \theta_1 - \theta_0 \le \Delta\}$: new treatment better than the control, but not by the required MID $\Delta$. 

In these numerical illustrations we choose the priors $\pi_e$ and  $\pi_f$ to be equal, both of the simple form of the product of two  Unif$(0,1)$-distributions. We denote this common prior by $\pi$. It is unlikely that one would want to use such a prior in any real application; for example, it has the same constant density $1$ near the  $(0,1)$ and  $(1,0)$ corners of the unit square, where the values of $\theta_1$ and $\theta_0$ are very far from each other, as close to the diagonal where they would be more similar. In a synthetic simulation experiment such as the present one this simple choice can perhaps be defended on the grounds that the posterior is now proportional to the product of two canonical Binomial likelihood expressions.

In these experiments,  data were generated in three different ways, corresponding to the alternatives (a), (b) and (c) above. To make this explicit, we denote by $\pi_{(a)}, \pi_{(b)}$ and $\pi_{(c)}$ three Uniform distributions for $(\theta_0,\theta_1)$ on the unit square, with $\pi_{(a)}$ being conditioned on configuration (a), $\pi_{(b)}$ on (b), and $\pi_{(c)}$ on (c). These distributions can be appropriately called  \textit{sampling priors}, e.g.,  \cite{lee2024using}. In each simulation, a pair of parameter values $(\theta_0,\theta_1)$ was drawn from the respective sampling prior, and outcome data $\{{\mathcal D}_{n};1 \le n \le N_{\max}\}$ were then generated accordingly. 

It is important to distinguish these three priors conceptually from $\pi$, which is used for computing the posterior probabilities in the analysis of the simulated data and which, therefore, can be appropriately called \textit{analysis prior}.

The remaining simulation parameters in the experiments were $\varepsilon_e = \varepsilon_f = 0.05$ and $\Delta = 0.05$, and the utility values $G_e = 2500,$ $G_f = 500,$ $L_e = L_f = 1000$.
The maximal trial size was $N_{max} = 500$, and the time horizon $T = 500.$    

Interim analyses were performed after each new observed outcome, i.e., $\sigma_n = n, 1 \le n \le 500$. Monte Carlo samples of size $1000$ were used in each step for evaluating the considered posterior probabilities and the predictive expected utilities.   Simulations of this latter type tend to be slow,  
as they involve sampling in the entire product space of the parameters $(\theta_0, \theta_1)$ and of possible data sequences $\{{\mathcal D}_{n};1 \le n \le 500\}$. Variance reduction
 and an alternative
stochastic filtering approach
are discussed in the Supplementary Material.

\begin{figure}[h!]
\captionsetup[subfigure]{labelformat=empty}
     \centering   
   \begin{subfigure}[t]{0.63\textwidth}
         \centering
         \includegraphics[width=\textwidth]{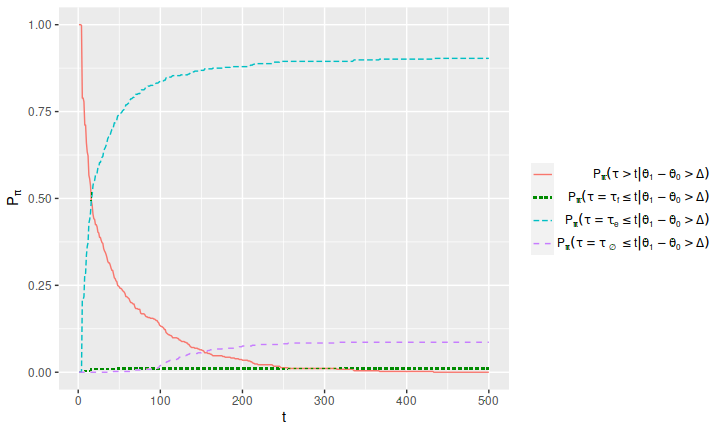}
     \end{subfigure}
     \hfill
     \begin{subfigure}[t]{0.63\textwidth}
         \centering
         \includegraphics[width=\textwidth]{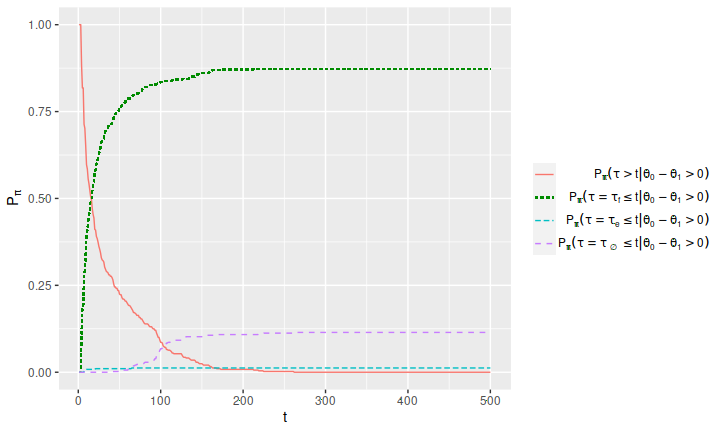}
     \end{subfigure}
     \hfill
\begin{subfigure}[t]{0.63\textwidth} 
         \centering
\includegraphics[width=\textwidth]{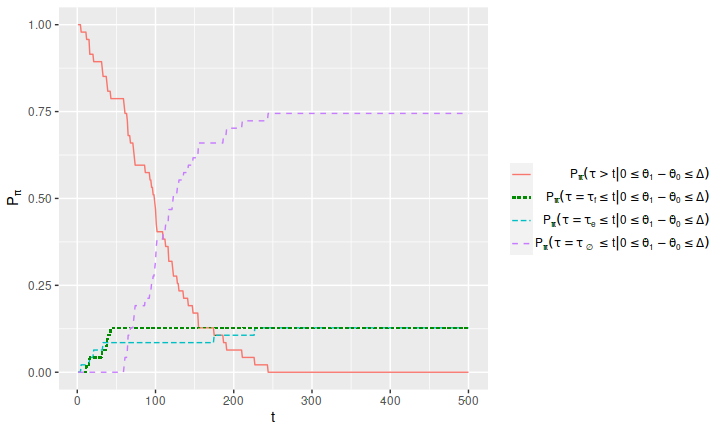}
\end{subfigure}
\caption{Predictive probabilities for different ways of stopping a trial by the time the outcomes from $t$ patients have been observed, subdivided according to the conclusion made: efficacy (\textit{turquoise}), futility (\textit{green}), or inconclusive (\textit{violet}. The curves in (\textit{red}) show the overall 'survival' probabilities of not having stopped the trial by time $t$ for any of these reasons.   In the top figure, the data were generated from sampling prior $\pi_{(a)}$ on   configuration  $\{\theta_1 - \theta_0 > \Delta\}$; thus  conclusion $d_e$ from the data analysis would be correct.  The \textit{turquoise} curve shows the values of the sub-CDF $t \rightarrow \P_{\pi}( \bftau = \bftau_e \le t \vert \theta_1 - \theta_0 > \Delta )$ for $0 \le t \le 500.$ 
In the middle figure, data were generated from   $\pi_{(b)}$ on     $\{\theta_1 - \theta_0 <0\}$; thus   $d_f$   would be correct, and the \textit{green} curve shows the sub-CDF $t \rightarrow \P_{\pi}( \bftau = \bftau_f \le t \vert \theta_1 - \theta_0 <0 )$. In the bottom figure, data were generated from   $\pi_{(c)}$ on     $\{0\le\theta_1 - \theta_0 \le \Delta\}$; thus   $d_{\oslash}$   would be correct, and the \textit{violet} curve shows the sub-CDF $t \rightarrow \P_{\pi}( \bftau = \bftau_{\oslash} \le t \vert \le\theta_1 - \theta_0 \le \Delta )$. Monte Carlo samples of size 1000 were used in each step for evaluating the considered posterior probabilities.
}
        \label{fig:1:2}
\end{figure}

\begin{figure}[h!] 
\captionsetup[subfigure]{labelformat=empty}
     \centering   
    \begin{subfigure}[t]{0.39\textwidth} 
         \centering
         \includegraphics[trim=1cm 0cm 0cm 0cm, clip=true,width=\textwidth]{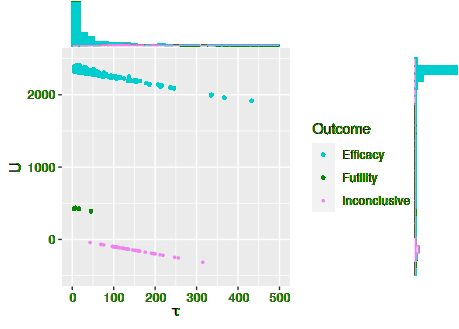} 
         \caption{ 
         $\E_{\pi}\bigl( 
         U_{\bftau}-\bftau \big\vert \bftheta_1 - \bftheta_0 > \Delta\bigr) = 2089.62
         $ }
\label{subfig:th1_>_th0_+_delta_onscale}
     \end{subfigure}
     ~
 \begin{subfigure}[t]{0.39\textwidth}
         \centering
         \includegraphics[trim=1cm 0cm 0cm 0cm, clip=true,width=\textwidth]{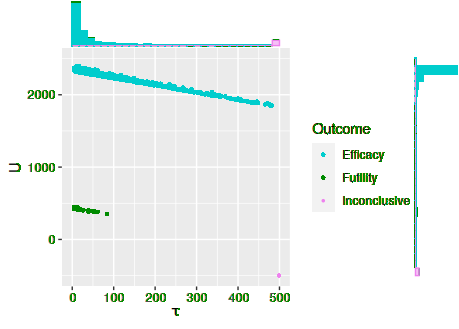}
         \caption{ 
         $
         \E_{\pi}\bigl( 
         U_{\bftau} -\bftau \big\vert 
         \bftheta_1 - \bftheta_0 > \Delta\bigr)=2093.77 $  }\label{subfig:th1_>_th0_+_delta_onscale_simple}
     \end{subfigure}
     \hfill 
     \begin{subfigure}[t]{0.39\textwidth}
         \centering
    \includegraphics[trim=1cm 0cm 0cm 0cm, clip=true,width=\textwidth]{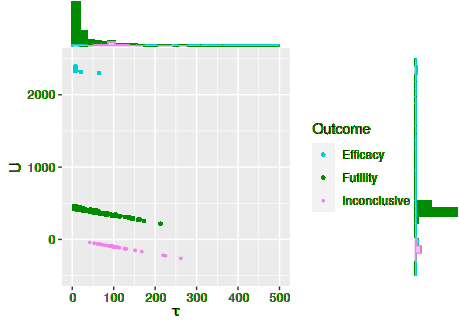}  
         \caption{   
         $\E_{\pi}\bigl(   U_{\bftau}-\bftau  
         \big\vert \bftheta_1 - \bftheta_0 < 0 \bigr)=377.70$
         }
         \label{fig:th0_>_th1}
     \end{subfigure}
     ~
    \begin{subfigure}[t]{0.39\textwidth}
         \centering
         \includegraphics[trim=1cm 0cm 0cm 0cm, clip=true,width=\textwidth]{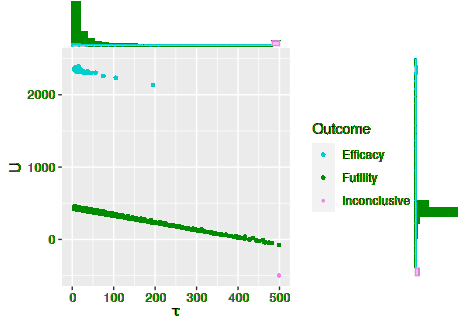} 
         \caption{  
         $\E_{\pi}\bigl( 
         U_{\bftau}-\bftau \big\vert
         \bftheta_1 - \bftheta_0 < 0
      \bigr)= 382.19   $      
         }         \label
         {fig:th0_>_th1_simple}
     \end{subfigure}   
     \hfill
     \begin{subfigure}[t]{0.39\textwidth}
         \centering
         \includegraphics[trim=1cm 0cm 0cm 0cm, clip=true,width=\textwidth]{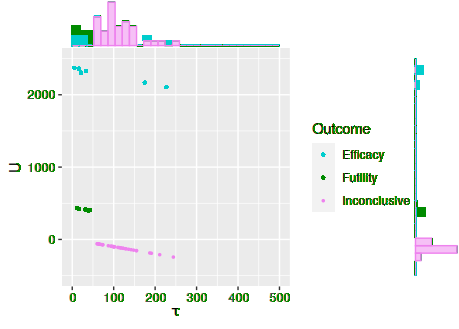}
         \caption{
         $\E_{\pi}\bigl(
         U_{\bftau}-\bftau\big\vert 
         0 \le \bftheta_1 - \bftheta_0 \le \Delta \bigr)= 257.66$}        \label{fig:th0_<_th1_<_th0_+_delta}
     \end{subfigure}
~          \begin{subfigure}[t]{0.39\textwidth}
         \centering
         \includegraphics[trim=1cm 0cm 0cm 0cm, clip=true,width=\textwidth]{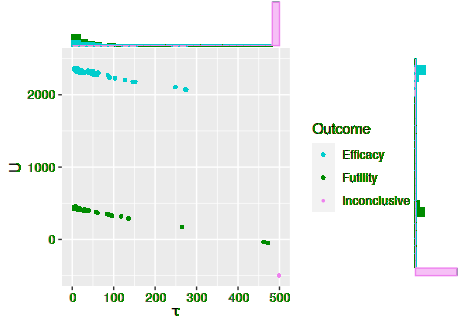}
         \caption{
 $        \E_{\pi}
         \bigl(
 U_{\bftau} -\bftau \big \vert    
 0 \le \bftheta_1 - \bftheta_0 \le \Delta \bigr)= 196.76$
 }
\label{fig:th0_<_th1_<_th0_+_delta_simple}
     \end{subfigure}        
\caption{An illustration of the effect of allowing  for the possibility of a   inconclusive 'early stopping' of the trial (left), and not  allowing for it (right). The scatter plots are formed by considering jointly the duration of the trial and the realized final conditionally expected utility value at the time the trial is stopped, by using the following color coding: $(\bftau_e, U_{\bftau_e}-\bftau_e)$ is \textit{turquoise}, $(\bftau_f, U_{\bftau_f}-\bftau_f)$  \textit{green}, and $(\bftau_{\oslash},  -\bftau_{\oslash})$  \textit{violet}. The marginal histograms of the durations and of the final utilities are coded similarly. In the top figure, the data were generated from sampling prior $\pi_{(a)}$ on   configuration  $\{\theta_1 - \theta_0 > \Delta\}$, in which case   conclusion $d_e$ from the data analysis would be correct.  
In the middle figure, the data were generated from   $\pi_{(b)}$ on     $\{\theta_1 - \theta_0 <0\}$, so that   $d_f$   would be correct. In the bottom figure, the data were generated from   $\pi_{(c)}$ on     $\{0\le\theta_1 - \theta_0 \le \Delta\}$, and   $d_{\oslash}$   would be correct. The expectations in each case are computed as sample averages. } 
\label{fig:joint_conditionals_under_alternatives}
\end{figure}

The results, shown in Figures \ref{fig:1:2} and \ref{fig:joint_conditionals_under_alternatives}, are intended as simple qualitative  illustrations of the ideas presented earlier, and particularly on how the probabilities of reaching each of the three decision alternatives $d_e, d_f$ and $d_{\oslash}$ depend on the choice of the  \textit{sampling prior}, here $\pi_{(a)}$, $\pi_{(b)}$ or $\pi_{(c)}$. 

The following crude conclusions can be drawn from the top part of Figure \ref{fig:1:2}: Here, the data were generated according to the sampling prior $\pi_{(a)}$ corresponding to  parameter configuration  $\{\theta_1 - \theta_0 > \Delta\}$; thus drawing the conclusion $d_e$ from the data analysis would be correct.  The \textit{turquoise} curve shows the values of the sub-CDF $t \rightarrow \P_{\pi}( \tau = \tau_e \le t \vert \theta_1 - \theta_0 > \Delta )$ for $0 \le t \le 500.$ It can be appropriately thought of as a power function for an efficacy conclusion as the number of observed treatment outcomes increases. Its values grow rapidly for small values of $t$, reach the level of approximately  $0.87$ at $t=200$, and finally about $0.90$ at $t=500$. The \textit{violet} curve $t \rightarrow \P_{\pi}( \tau = \tau_{\oslash} \le t \vert \theta_1 - \theta_0 > \Delta )$ corresponding to the inconclusive decision $d_{\oslash}$ remains very low for approximately the first $100$ values of $t$, but then increases to about 0.10 at $t= 300$, staying nearly constant thereafter. The \textit{green} curve $t \rightarrow \P_{\pi}( \tau = \tau_f\le t \vert \theta_1 - \theta_0 > \Delta )$ corresponding to concluded futility $d_f$, here a false negative, remains very low for all $t$. 

In the middle part of Figure \ref{fig:1:2}, with data  generated according to sampling prior  $\pi_{(b)}$ conditioned on   configuration  $\{\theta_1 - \theta_0< 0\}$, the probabilities of (correctly) concluding futility $d_f$ by time $t$ dominate those of $d_e$ and $d_{\oslash} $, with the probabilities of $ d_{\oslash}$ being slightly higher than in alternative (a).

In the bottom part of    Figure \ref{fig:1:2}, with the data coming from sampling prior  $\pi_{(c)}$ and thereby from parameter configuration  $\{0 \le\theta_1 - \theta_0 \le \Delta\}$, the  probabilities of (correctly) concluding $ d_{\oslash}$  by time $t$ dominate the alternatives $d_e$ and $d_f$, with the corresponding \textit{violet} curve reaching approximately the level $0.75$ at $t=250$. Note that, again, this possibility of 'early stopping' does not become active before approximately $60$ outcomes have been observed. Looking at the \textit{green} and \textit{turquoise} curves corresponding to the (incorrect) conclusions $d_f$ ans $d_e$, it appears to be somewhat more likely to come, early on, to the former than the latter. This, apparently, is due to that the  assumed uniform \textit{analysis prior} $\pi$ has more mass on parameter configuration   $\{  \theta_1-\theta_0<0\}$ than on $\{ \theta_1-\theta_0 >\Delta \}$.  

Figure \ref{fig:joint_conditionals_under_alternatives}  makes a comparison  between two  designs: one that allows for the possibility of a   inconclusive 'early stopping' of the trial (left), and one that does not (right), by considering jointly the duration $\tau$ of the trial and the realized final expected utility value $U_{\tau}$ at the time the trial is stopped. Each  colored dot of the scatter plots  corresponds to the outcome of a simulated trial, coded with colors as follows: $(\tau_e, U_{\tau_e}-\tau_e)$ is \textit{turquoise}, $(\tau_f, U_{\tau_f}-\tau_f)$  \textit{green}, and $(\tau_{\oslash},  -\tau_{\oslash})$  \textit{violet}. Here, $U_{\tau_e}$
refers here to  expression \eqref{eq:no:3} and $U_{\tau_f}$
to expression \eqref{eq:no:4}. The expected utilities are computed as averages across all simulations.

The two top figures correspond again to the situation where the data are generated according to sampling prior $\pi_{(a)}$, the middle figures according to  $\pi_{(b)}$, and the bottom ones according to $\pi_{(c)}$. In the case of  $\pi_{(a)}$ and  $\pi_{(b)}$, for which the respective correct conclusions would be $d_e$ and $d_f$, non-availability of the early stopping option has the effect that the trial can continue for much longer,  sometimes until the maximal trial size $N_{max}=500$. However, in both cases, the  difference of the conditional expected utilities computed across all simulated trials, between left and right, is small. This is because, in both situations, as demonstrated in the top and middle graphs of Figure \ref{fig:1:2}, the probability of arriving at the correct decision grows rapidly with the number observed outcomes, and thereby the need to consider the option of the inconclusive decision $d_{\oslash}$ remains marginal. 
The result depends naturally on the selected threshold values $\varepsilon_e$ and $\varepsilon_f$ controlling the possibilities of making a false conclusion, and on how the utility values $G_e, G_f,L_e$ and $L_f$ relate to the assumed unit cost of continuing the trial by one more patient: increasing the utility values makes it more attractive to run the trial for longer, and will thereby, with everything else fixed,  (stochastically) increase the value of $\tau_{\oslash}$

In the bottom part of Figure \ref{fig:joint_conditionals_under_alternatives}, where the data come from parameter configuration $\{0 \le \theta_1 - \theta_0 \le \Delta\}$, the desired result from the trial would be inconclusive stopping $d_{\oslash}$. When the design includes the option of making such a decision 'early' (left),  these conclusions become more likely and start to dominate the alternatives $d_e$ and $d_f$ after  approximately  $t=50$ outcomes  have been observed. In contrast, when the option of early stopping is removed (right),  a large proportion of the simulated  trials end only after the maximal size $N_{max}=500$ has been reached. The cost  that is incurred by the longer running times is  then reflected in the lower value of the expected utility.   

\section{Discussion}
\label{section:no:5}
 
Clinical trials are an instrument for making informed decisions based on experimental data. In phase 2 trials, the usual goal is to make a comparative evaluation on the efficacy of an experimental treatment to  control, and in multi-arm trials, also to each other. More successful treatments among the  alternatives  considered, if found, can then be selected for a confirmatory analysis in phase 3.  Essentially the same ideas and methods can  be used  in both phase 2 and phase 3, and even in a seamless fashion unless it is felt that the change from an exploratory to a confirmatory mode of analysis should be reflected in selecting a tighter bound $\varepsilon_e$ and/or a more conservative prior.
An additional possibility (cf. \cite{nikolakopoulos2016hybrid}) is to use the method of Section \ref{subsection:no:2.2} as an aid, after having successfully completed phase 2, to decide whether it is worthwhile to go on and proceed further to phase 3.

Such conclusions should be made using relevant prior information and the data coming from the trial itself, with the prior being based on trusted expert assessments and, whenever possible, on existing empirical data from comparable treatments. This is precisely what the consequent Bayesian posterior distributions synthesize and express. Compelling arguments justifying their use in the context of clinical trials have been presented in the literature since  \cite{berry1985interim}. The "naturally dynamic" character of Bayesian trial designs, including the computation of predictive power in interim analyses, 
was presented in the papers
\cite{SPIEGELHALTER19868} and \cite{spiegelhalter1994bayesian}. In view of this,  the present paper is not claimed to contain great  novelty of ideas, and instead be seen as a commentary on the current, sometimes problematic practice of applying Bayesian trial designs, now forty years from  the appearance of the seminal contributions. 
For an interesting more recent opinion, see the blog by \citet*{https://www.fharrell.com/post/bayes-seq/}, which  contains also a demonstration of how decisions based on stopping boundaries for Bayesian posteriors are well calibrated. 

Of course, such views cannot be enforced on unwilling members of the clinical trials community. An expert in the area, after having read an early version of our text, wrote as follows:
"My main concern is that the added value of the suggested approach is in no way shown, whereas strong claims are made in favor of Bayesian approaches. To substantiate this, operating characteristics need to be calculated for various true values of the outcome parameters under both treatments, including expected sample size,   probability of early stopping for efficacy, rejection rates (type I error rate or power), false discovery probabilities, true discovery probabilities and probability of a correct decision. These should be compared to frequentist trials (incorporating more traditional interim analyses). This is needed to illustrate whether calculations  are feasible, to show how the suggested approach behaves in real-life applications and to show whether control of FDP as proposed results in more efficient trials (as implicitly claimed by the authors) without for instance jeopardizing probability of correct decisions." 

According to this expert, a large-scale research program based on both computer simulations and real-life data, should be carried out before presentation of opinions dissenting from the dominant NHST-driven paradigm would be allowed. This is despite that there is increasing concern that the yardstick provided by the NHST paradigm in its commonly practiced form is itself seriously deficient, and that this is an important contributor to the prevailing replication crisis in experimental, including bio-pharmaceutical, research. Worth reading are, for example, the early paper  \cite{rozeboom1960fallacy}, the influential work \cite{ioannidis2005most}, and \cite{doi:10.1080/00031305.2016.1154108} relating to the ASA \textit{Statement on p-Values and Statistical Significance} (2016).

Second, computer experiments, however extensive, cannot be used for resolving foundational differences between statistical paradigms. Our proposal for error control, like all Bayesian statistics, rests on assumed validity of the  conditionality principle: the results from data analysis are expressed in terms of posterior probabilities, conditioned on the data that were actually observed, and in here probabilities are viewed as expressions of epistemic uncertainty concerning the unknowns, including the values of the model parameters.   Frequentist methods for error control, in contrast,  take into account outcomes that could have occurred but did not, such as might be observed from hypothetical trials performed under similar conditions in the future, and altogether preclude assigning probabilities to statements concerning parameter values.  

It needs to be emphasized that, while this interpretation of posterior probabilities as expressions of uncertainty can appear "soft" when compared to the limiting relative frequency interpretation customarily used in frequentist statistics,  their values, as conditional probabilities given the data, are fully determined by probability calculus from the   ingredients \{prior, likelihood, data\}. In the present context, the likelihood has canonical Binomial form, and the approval of the prior $\pi_e$ is assumed to be at the discretion of the regulators.

Every trial design contains a plan for the procedure according to which patients will be assigned to the  treatments, how the outcome data will be analyzed, and on how the conclusions from such analyses are to be drawn. We have considered the latter two aspects. Due to the assumed likelihood principle, the conclusions from the data analysis do not depend on the form of the assignment mechanism if  only the consequent likelihood contribution 
does not depend on the targeted parameters, here $\theta_0$ and $\theta_1$.  For example, the well-known \textit{Response Adaptive Randomization} (RAR) scheme satisfies this condition if the randomization is performed externally.  

The decision rules   considered here for superiority trials are easily modified to be appropriate for \textit{equivalence} and \textit{non-inferiority} trials as well. For example, for the latter, we would be led to considering posterior probabilities of the form $ \P_{\pi_e}  (  \bftheta _{1} - \bftheta _{0} \geq -\Delta \vert {\mathcal D}_{\sigma_n}  )$, where $\Delta$ is a selected non-inferiority margin. Moreover, the prototype case of dichotomous outcomes measured soon after treatment   can be modified to other types of outcome data without changing the logical basis of the method if the relevant posterior probabilities are computable  from the accruing data. Normally distributed outcomes, with known variance, are particularly easy to deal with when associated with Normal priors. Extensions into sequential multi-arm designs, and thereby applying adaptive rules for treatment assignment, are also feasible and will be considered in later work. 

Proper utilization of Bayesian decision theory for drawing the conclusions would involve, from the very beginning, incorporation of utility functions into a corresponding model (e.g.,  \cite{rosner2020bayesian}).
We have made a shortcut in this regard, by first only considering, in Section \ref{section:no:2}, the threshold values $\varepsilon_e$ and $\varepsilon_f$ for concluding $d_e$ or $d_f$, and thereby deferring utility considerations to Section \ref{subsection:no:2.2}, where we included the option $d_{\oslash}$ of stopping the trial early. Even then, we have not tackled actual optimization issues.
With this, we wanted to facilitate the practical implementation of such designs. Specification of appropriate values for  $\varepsilon_e$ and $\varepsilon_f$ should be relatively easy, due to their intuitively straightforward probability interpretation.

Elicitation of values for the utilities $G_e$, $G_f$, $L_e$ and $L_f$, although  necessary  for a fully satisfactory solution of the problem of inconclusive early stopping, requires more careful thought on the part of the domain specialists and the different stakeholders in question. An increase in the value of any of these utilities has the effect of shifting $\tau_{\oslash}$ further in time. 

A central issue in this paper has been our attempt to find a satisfactory answer to the question: If the statistical analysis of trial data is done by applying the tools from Bayesian inference and decision theory, how should one respond to the regulators' guidelines on applying standard frequentist methods for error control?

We have been critical towards the \textit{Frequentist-oriented Perspective}, as described by \cite{zhou_ji_2023}, for reasons that are both conceptual and practical. It is a hybrid approach which both relies on the likelihood principle in Bayesian estimation and violates it in frequentist error control.  For computing the value of type I error rate
or of the more general \textit{Family-wise Error Rate} (FWER) 
one needs to account  for the consequent multiplicity problem in testing, and thereby for 
all possible ways of making errors in concluding efficacy that may turn up when the trial is to be run. This task becomes more burdensome in multi-arm designs, increasingly so in Multi-Arm Multi-Stage (MAMS) designs, and is literally infeasible in platform designs, where some prospective treatment groups are not even known at the time the master protocol is written.  

Systematic sequential monitoring of the outcomes from the trial, and making corresponding interim checks, may be avoided in practice for valid reasons such as greater complexity in the required logistics, more elaborate computations, or because it is argued that the consequent unblinding could jeopardize the whole study. However, looking from a Bayesian perspective, avoiding interim checks because this would inflate type I error rate, as would be the case when employing the technical machinery of $\alpha$-spending functions, is an artifact stemming from methods that do not respect the conditionality principle.
Additional insightful comments on the implications of applying the likelihood principle in the clinical trials context can be found in \cite{https://www.fharrell.com/post/pvalprobs/}.

In our discussion, in \ref{subsection:no:3.2}, of the \textit{Calibrated Bayesian Perspective} we argued that in the analysis of trial data frequentist methods would find their natural role in helping the elicitation of the prior, and again, in phase 4, in monitoring real-world data and then verifying that the results are in harmony with the conclusions and predictions made earlier. Thus, our main criticism is not directed against applying frequentist methods  in general, but against using   hybrids of the two mutually conflicting statistical paradigms for the same selected purpose. Merging them in a single trial design only leads to a difficult-to-understand mongrel of a method. Or, to put it differently, one should choose whether to use a belt or suspenders. One is enough, and better than a combination of the two. Calibration, by loosening or tightening one as determined by  the other, only confuses the  natural function of the selected means.
Thus, our answer to the question in the title of this contribution is a 
definite 'No'.

Currently the regulators rarely receive   requests to handle Bayesian trial designs for possible approval (e.g., \cite{medical2022not}).  
Exceptions include, on the one hand, well-known instances of multicenter platform trials such as I-SPY2 (e.g., \cite{wang2019spy}), which uses Bayesian adaptive randomization to match patients to drugs based on molecular biomarkers, and on the other, small size  
trials during the early phases of drug development and testing, often
applying basket or umbrella designs and using  hierarchical Bayesian models  for the purpose of "borrowing strength" across different strata with only a small number of patients in each. 

Despite  applying Bayesian ideas in such novel ways, these designs are commonly furnished with a frequentist method for error control,  
fortified by extensive preliminary simulation experiments to determine traditional operating characteristics for a variety of different design parameter values. The reason may be
methodological conservatism, but perhaps also lacking familiarity with the Bayesian core principles, and even prejudices towards such methods, based on superficially understood notions of 'subjective' and 'objective' in science (e.g., \cite{berger1988statistical}, \cite{draper2006coherence})"The difficulty lies not so much in developing new ideas as in escaping from old ones" (J. M. Keynes).

An important recent development on the regulatory front is the appearance of   \cite{fda_190505}, a draft guidance  on the use of the Bayesian methodology in clinical trials. It is very cautious in its recommendations but, in particular, no longer insists that the trials need be controlled for type I error rate. Two  commentaries  written by well known authorities in the area, \cite{ 10.1001/jama.2026.4179} and \cite{10.1001/jama.2026.4175}, were published soon after the \cite{fda_190505} document was made available. Between them, opinions on applying the Bayesian methodology in practical trial designs were as divided as ever before.

\printbibliography

@article{arjas2016wearing,
  title={How About Wearing Two Hats, First Popper's and then de Finetti's?},
  author={Arjas, Elja},
  journal={Statistical Science},
  volume={31},
  number={4},
  pages={545--548},
  year={2016},
  publisher={JSTOR}
}

@article{Arjas2022,
  doi  =  {10.1186/s12874-022-01526-8},
  url  =  {https://doi.org/10.1186/s12874-022-01526-8},
  year  =  {2022},
  month  =  {2},
  publisher  =  {Springer Science and Business Media {LLC}},
  volume  =  {22},
  number  =  {1},
  author  =  {Elja Arjas and Dario Gasbarra},
  title  =  {Adaptive treatment allocation and selection in multi-arm clinical trials: a Bayesian perspective},
  journal  =  {{BMC} Med Res Methodol}
}

@article{azzolina2021prior,
  title={Prior elicitation for use in clinical trial design and analysis: a literature review},
  author={Azzolina, Danila and Berchialla, Paola and Gregori, Dario and Baldi, Ileana},
  journal={International journal of environmental research and public health},
  volume={18},
  number={4},
  pages={1833},
  year={2021},
  publisher={MDPI}
}

@article{bassi2021bayesian,
  title  =  {Bayesian adaptive decision-theoretic designs for multi-arm multi-stage clinical trials},
  author  =  {Bassi, Andrea and Berkhof, Johannes and de Jong, Daphne and van de Ven, Peter M},
  journal  =  {Stat Methods Med Res},
  volume  =  {30},
  number  =  {3},
  pages  =  {717--730},
  year  =  {2021},
  publisher  =  {SAGE Publications Sage UK: London, England}
}

@article{Beall2022InterpretingAB,
  title  =  {Interpreting a Bayesian phase II futility clinical trial},
  author  =  {Jonathan Beall and Christy N Cassarly and Renee L Martin},
  journal  =  {Trials},
  year  =  {2022},
  volume  =  {23}
}

@article{benjamini_1995,
author  =  {Benjamini, Yoav and Hochberg, Yosef},
title  =  {Controlling the False Discovery Rate: A Practical and Powerful Approach to Multiple Testing},
journal  =  { J R Stat Soc Series B Stat Methodol},
volume  =  {57},
number  =  {1},
pages  =  {289-300},
doi  =  {https://doi.org/10.1111/j.2517-6161.1995.tb02031.x},
url  =  {https://rss.onlinelibrary.wiley.com/doi/abs/10.1111/j.2517-6161.1995.tb02031.x},
year  =  {1995}
}

@article{berger1988statistical,
  title={Statistical analysis and the illusion of objectivity},
  author={Berger, James O and Berry, Donald A},
  journal={American scientist},
  volume={76},
  number={2},
  pages={159--165},
  year={1988},
  publisher={JSTOR}
}

@book{berger,
    AUTHOR  =  {Berger, James O. and Wolpert, Robert L.},
     TITLE  =  {The likelihood principle},
    SERIES  =  {Institute of Mathematical Statistics Lecture Notes---Monograph
              Series},
    VOLUME  =  {6},
 PUBLISHER  =  {Institute of Mathematical Statistics, Hayward, CA},
      YEAR  =  {1984},
     PAGES  =  {xi+206},
      ISBN  =  {0-940600-06-4},
   MRCLASS  =  {62A15 (62A20)}
}

@article{berry1985interim,
  title  =  {Interim analyses in clinical trials: classical vs. Bayesian approaches},
  author  =  {Berry, Donald A.},
  journal  =  {Stat Med},
  volume  =  {4},
  number  =  {4},
  pages  =  {521--526},
  year  =  {1985},
  publisher  =  {Wiley Online Library}
}

@article{BERRY1999215,
title = {Bayesian perspectives on multiple comparisons},
journal = {Journal of Statistical Planning and Inference},
volume = {82},
number = {1},
pages = {215-227},
year = {1999},
issn = {0378-3758},
doi = {https://doi.org/10.1016/S0378-3758(99)00044-0},
url = {https://www.sciencedirect.com/science/article/pii/S0378375899000440},
author = {Donald A Berry and Yosef Hochberg}
}

@book{berry,
    AUTHOR  =  {Berry, Scott M. and Carlin, Bradley P. and Lee, J. Jack and
              M\"{u}ller, Peter},
     TITLE  =  {Bayesian adaptive methods for clinical trials},
    SERIES  =  {Chapman \& Hall/CRC Biostatistics Series},
    VOLUME  =  {38},
      NOTE  =  {With a foreword by David J. Spiegelhalter},
 PUBLISHER  =  {CRC Press},
      YEAR  =  {2011},
     PAGES  =  {xviii+305},
      ISBN  =  {978-1-4398-2548-8},
   MRCLASS  =  {62P10 (62C10 62F15 92C50)},
  MRNUMBER  =  {2723582}
}

@book{buck2006uncertain,
  title={Uncertain Judgements: Eliciting Experts' Probabilities. Statistics in Practice},
  author={Buck, Caitlin E and Daneshkhah, Alireza},
  year={2006},
  publisher={Wiley}
}

@article{carlin1998approaches,
  title = {Approaches for optimal sequential decision analysis in clinical trials},
  author = {Carlin, Bradley P and Kadane, Joseph B and Gelfand, Alan E},
  journal = {Biometrics},
  pages = {964--975},
  year = {1998},
  publisher = {JSTOR}
}

@article{cheng2005bayesian,
  title = {Bayesian adaptive designs for clinical trials},
  author = {Cheng, Yi and Shen, Yu},
  journal = {Biometrika},
  volume = {92},
  number = {3},
  pages = {633--646},
  year = {2005},
  publisher = {Oxford University Press}
}

@article{chevret2012bayesian,
  title = {Bayesian adaptive clinical trials: a dream for statisticians only?},
  author = {Chevret, Sylvie},
  journal = {Stat Med},
  volume = {31},
  number = {11-12},
  pages = {1002--1013},
  year = {2012},
  publisher = {Wiley Online Library}
}

@article{dallow2011perils,
  title = {The perils with the misuse of predictive power},
  author = {Dallow, Nigel and Fina, Paolo},
  journal = {Pharm Stat},
  volume = {10},
  number = {4},
  pages = {311--317},
  year = {2011},
  publisher = {Wiley Online Library}
}

@article{dallow2018better,
  title={Better decision making in drug development through adoption of formal prior elicitation},
  author={Dallow, Nigel and Best, Nicky and Montague, Timothy H},  journal={Pharmaceutical Statistics},
  volume={17},
  number={4},
  pages={301--316},
  year={2018},
  publisher={Wiley Online Library}
}

@article{SangitaIsha2023,
title  =  "Weighted U-statistics for likelihood-ratio ordering of bivariate data",
author  =  "Sangita Kulathinal and Isha Dewan",
year  =  "2023",
month  =  apr,
doi  =  "10.1007/s00362-022-01332-w",
language  =  "English",
volume  =  "64",
pages  =  "705--735",
journal  =  "Statistical Papers",
issn  =  "0932-5026",
publisher  =  "Springer",
number  =  "2",
}

@article{diaconis2023approximate,
  title={Approximate exchangeability and de Finetti priors in 2022},
  author={Diaconis, Persi},
  journal={Scandinavian Journal of Statistics},
  volume={50},
  number={1},
  pages={38--53},
  year={2023},
  publisher={Wiley Online Library}
}

@article{dmitrienko2006bayesian,
  title = {Bayesian predictive approach to interim monitoring in clinical trials},
  author = {Dmitrienko, Alexei and Wang, Ming-Dauh},
  journal = {Stat Med},
  volume = {25},
  number = {13},
  pages = {2178--2195},
  year = {2006},
  publisher = {Wiley Online Library}
}

@article{draper2006coherence,
  title = {Coherence and calibration: comments on subjectivity and "objectivity" in Bayesian analysis (comment on articles by Berger and by Goldstein)},
  author = {Draper, David},
journal = {Bayesian Anal},
  volume = {1},
  number = {3},
  pages = {423--428},
  year = {2006}
}

@article{efron1986isn,
  title={Why isn't everyone a Bayesian?},
  author={Efron, Bradley},
  journal={The American Statistician},
  volume={40},
  number={1},
  pages={1--5},
  year={1986},
  publisher={Taylor \& Francis}
}

@misc{EMA_CHMP_44762_2017,
author  =  {{EMA}},
title  =  {Guideline on multiplicity issues in clinical trials},
URL ={https://www.ema.europa.eu/en/documents/scientific-guideline/draft-guideline-multiplicity-issues-clinical-trials_en.pdf/},
year  =  {2017}
}

@article{10.1001/jama.2026.4175,
    author = {Evans, Scott R. and Fleming, Thomas R. and Janes, Holly and Dodd, Lori E.},
    title = {Reflections on FDA Draft Guidance on Bayesian Methods in Trials—Protecting Scientific Integrity and Evidentiary Standards},
    journal = {JAMA},
    volume = {335},
    number = {19},
    pages = {1659-1661},
    year = {2026},
    month = {05},
    issn = {0098-7484},
    doi = {10.1001/jama.2026.4175},
    url = {https://doi.org/10.1001/jama.2026.4175},
    eprint = {https://jamanetwork.com/journals/jama/articlepdf/2847013/jama_evans_2026_pp_260010_1778177984.35458.pdf},
}

@article{fayers1997tutorial,
  title={Tutorial in biostatistics: Bayesian data monitoring in clinical trials},
  author={Fayers, Peter M and Ashby, Deborah and Parmar, Mahesh KB},
  journal={Statistics in medicine},
  volume={16},
  number={12},
  pages={1413--1430},
  year={1997},
  publisher={Wiley Online Library}
}

@misc{ICH_E20,
author = {{ICH}},
title = {E20 Guideline on adaptive designs for clinical trials, Draft version.},
URL = {https://www.ema.europa.eu/en/documents/scientific-guideline/ich-e20-guideline-adaptive-designs-clinical-trials-step-2b_en.pdf},
year = {2025}
}

@misc{fda_71512,
author  =  {{FDA}},
title  =  {Guidance for the Use of Bayesian Statistics in Medical Device Clinical Trials.},
URL = {https://www.fda.gov/media/71512/download/},
year  =  {2010}
}

@misc{fda_78495,
author  =  {{FDA}},
title  =  {Adaptive Designs for Clinical Trials of Drugs and Biologics: Guidance for Industry.},
URL = {https://www.fda.gov/media/78495/download/},
year  =  {2019}
}

@misc{fda_190505,
author  =  {{FDA}},
title  =  { Use of Bayesian
Methodology in
Clinical Trials of Drug
and Biological Products
Guidance for Industry,
DRAFT GUIDANCE.},
URL = {https://www.fda.gov/media/190505/download/},
year  =  {2026}
}

@article{freedman1994and,
  title = {The what, why and how of Bayesian clinical trials monitoring},
  author = {Freedman, Laurence S and Spiegelhalter, David J and Parmar, Mahesh KB},
  journal = {Stat Med},
  volume = {13},
  number = {13-14},
  pages = {1371--1383},
  year = {1994},
  publisher = {Wiley Online Library}
}

@article{geisser1994interim,
  title = {Interim analysis for normally distributed observables},
  author = {Geisser, Seymour and Johnson, Wesley},
  journal = {Lect Notes Monogr Ser},
  pages = {263--279},
  year = {1994},
  publisher = {JSTOR}
}

@Article{ijerph18020530,
AUTHOR  =  {Giovagnoli, Alessandra},
TITLE  =  {The Bayesian Design of Adaptive Clinical Trials},
JOURNAL  =  {Int J Environ Res Public Health},
VOLUME  =  {18},
YEAR  =  {2021},
NUMBER  =  {2},
ARTICLE-NUMBER  =  {530},
URL  =  {https://www.mdpi.com/1660-4601/18/2/530},
PubMedID  =  {33435249},
ISSN  =  {1660-4601},
DOI  =  {10.3390/ijerph18020530}
}

@book{gradshteyn,
    author  =  {Gradshteyn, I. S. and Ryzhik, I. M.},
    pages  =  {xlviii+1171},
  publisher  =  {Elsevier/Academic Press},
  title  =  {Table of integrals, series, and products},
  year  =  {2007}
}

@article{Grieve2016IdleTO,
  title = {Idle thoughts of a 'well-calibrated' Bayesian in clinical drug development.},
  author = {Andrew P. Grieve},
  journal = {Pharmaceutical statistics},
  year = {2016},
  volume = {15 2},
  pages = { 96-108 }
}

@article{https://www.fharrell.com/post/bayes-seq/,
author  =  {Harrell, Frank},
title  =  {Continuous Learning from Data: No Multiplicities from Computing and Using Bayesian Posterior Probabilities as Often as Desired},
journal  =  {Statistical Thinking, Blog},
year  =  {2017},
month = { 10 },
url = {https://www.fharrell.com/post/bayes-seq/}
}

@article{https://www.fharrell.com/post/pvalprobs/,
author  =  {Harrell, Frank},
title  =  {p-values and Type I Errors are Not the Probabilities We Need},
journal  =  {Statistical Thinking, Blog},
year  =  {2017},
month = { 1 },
url = {https://www.fharrell.com/post/pvalprobs/}
}

@article{Herson1979PredictivePE,
  title = {Predictive probability early termination plans for phase II clinical trials.},
  author = {J. H. Herson},
  journal = {Biometrics},
  year = {1979},
  volume = {35 4},
  pages = { 775-83 }
}

@article{ioannidis2005most,
  title={Why most published research findings are false},
  author={Ioannidis, John PA},
  journal={PLoS medicine},
  volume={2},
  number={8},
  pages={e124},
  year={2005},
  publisher={Public Library of Science}
}

@article{johns1999use,
  title = {Use of predictive probabilities in phase II and phase III clinical trials},
  author = {Johns, Don and Andersen, John S},
  journal = {J Biopharm Stat },
  volume = {9},
  number = {1},
  pages = {67--79},
  year = {1999},
  publisher = {Taylor \& Francis}
}

@article{kass1989investigating,
  title={[Investigating Therapies of Potentially Great Benefit: ECMO]: Comment},
  author={Kass, Robert E and Greenhouse, Joel B},
  journal={Statistical Science},
  volume={4},
  number={4},
  pages={310--317},
  year={1989},
  publisher={JSTOR}
}

@article{kopp2019monitoring,
  title={Monitoring futility and efficacy in phase II trials with Bayesian posterior distributions—A calibration approach},
  author={Kopp-Schneider, Annette and Wiesenfarth, Manuel and Witt, Ruth and Edelmann, Dominic and Witt, Olaf and Abel, Ulrich},
  journal={Biometrical Journal},
  volume={61},
  number={3},
  pages={488--502},
  year={2019},
  publisher={Wiley Online Library}
}

@article{Lee2012BayesianCT,
  title = {Bayesian clinical trials in action.},
  author = {J. Jack Lee and Caleb T. Chu},
  journal = {Stat Med},
  year = {2012},
  volume = {31 25},
  pages = {
          2955-72
        }
}

@article{lee2008predictive,
  title = {A predictive probability design for phase II cancer clinical trials},
  author = {Lee, J Jack and Liu, Diane D},
  journal = {Clin Trials},
  volume = {5},
  number = {2},
  pages = {93--106},
  year = {2008},
  publisher = {Sage Publications Sage UK: London, England}
}

@article{lee2024using,
  title={Using Bayesian statistics in confirmatory clinical trials in the regulatory setting: a tutorial review},
  author={Lee, Se Yoon},
  journal={BMC Medical Research Methodology},
  volume={24},
  number={1},
  pages={110},
  year={2024},
  publisher={Springer}
}

@article{10.1001/jama.2026.4179,
    author = {Lee, J. Jack and Harrell, Frank E., Jr and LaVange, Lisa M. and Spiegelhalter, David J.},
    title = {Embracing Bayesian Methods in Clinical Trials: FDA’s Long-Awaited Draft Guidance},
    journal = {JAMA},
    volume = {335},
    number = {19},
    pages = {1664-1666},
    year = {2026},
    month = {05},
    issn = {0098-7484},
    doi = {10.1001/jama.2026.4179},
    url = {https://doi.org/10.1001/jama.2026.4179},
    eprint = {https://jamanetwork.com/journals/jama/articlepdf/2847011/jama_lee_2026_pp_260012_1778177982.76499.pdf},
}

@book{lesaffre2020,
  title = {Bayesian Methods in Pharmaceutical Research },
  author = { Lesaffre, Emmanuel and Baio, Gianluca and Boulanger, Bruno},
  ISBN = {9781315180212},
  url = {http://dx.doi.org/10.1201/9781315180212},
  DOI = {10.1201/9781315180212},
  month = { 4 }, 
  editor = {Lesaffre, Emmanuel and Baio, Gianluca and Boulanger, Bruno},
  year = {2020},
  publisher = { CRC Press}
}

@article{lin2020novel,
  title = {Novel bayesian adaptive designs and their applications in cancer clinical trials},
  author = {Lin, Ruitao and Lee, J Jack},
  journal = {Computational and Methodological Statistics and Biostatistics },
  pages = {395--426},
  year = {2020},
  publisher = {Springer}
}

@article{doi:10.1198/000313006X117837,
author  =  {Roderick J Little},
title  =  {Calibrated Bayes},
journal  =  {Am Stat},
volume  =  {60},
number  =  {3},
pages  =  {213-223},
year  =  {2006},
publisher  =  {Taylor \& Francis},
doi  =  {10.1198/000313006X117837},
URL  =  {
        https://doi.org/10.1198/000313006X117837
},
 
}

@book{Marshall2011,
  title = {Inequalities: Theory of Majorization and Its Applications},
  author = {Marshall,  Albert W. and Olkin,  Ingram and Arnold,  Barry C.},
  ISBN = {9780387682761},
  ISSN = {2197-568X},
  url = {http://dx.doi.org/10.1007/978-0-387-68276-1},
  DOI = {10.1007/978-0-387-68276-1},
  journal = {Springer Series in Statistics},
  publisher = {Springer},
  year = {2011}
}

@article{article,
author  =  {Mayo, Matthew and Gajewski, Byron},
year  =  {2004},
month  =  {05},
pages  =  {157-67},
title  =  {Bayesian sample size calculations in phase II clinical trials using informative conjugate priors},
volume  =  {25},
journal  =  {Controlled clinical trials},
doi  =  {10.1016/j.cct.2003.11.006}
}

@article{medical2022not,
  title={Why are not there more Bayesian clinical trials? Perceived barriers and educational preferences among medical researchers involved in drug development},
  author={ Clark, Jennifer and Muhlemann, Natalia and Natanegara, Fanni and Hartley, Andrew and Wenkert, Deborah and Wang, Fei and Harrell, Frank E and Bray, Ross},
  journal={Therapeutic Innovation \& Regulatory Science},
  pages={1--9},
  year={2022},
  publisher={Springer}
}

@article{mikkola2024prior,
  title={Prior knowledge elicitation: The past, present, and future},
  author={Mikkola, Petrus and Martin, Osvaldo A and Chandramouli, Suyog and Hartmann, Marcelo and Abril Pla, Oriol and Thomas, Owen and Pesonen, Henri and Corander, Jukka and Vehtari, Aki and Kaski, Samuel and others},
  journal={Bayesian Analysis},
  volume={19},
  number={4},
  pages={1129--1161},
  year={2024},
  publisher={International Society for Bayesian Analysis}
}

@article{morita2010evaluating,
  title={Evaluating the impact of prior assumptions in Bayesian biostatistics},
  author={Morita, Satoshi and Thall, Peter F and M{\"u}ller, Peter},
  journal={Statistics in biosciences},
  volume={2},
  pages={1--17},
  year={2010},
  publisher={Springer}
}

@article{muehlemann2023tutorial,
  title = {A Tutorial on Modern Bayesian Methods in Clinical Trials},
  author = {Muehlemann, Natalia and Zhou, Tianjian and Mukherjee, Rajat and Hossain, Munshi Imran and Roychoudhury, Satrajit and Russek-Cohen, Estelle},
  journal = {Therapeutic Innovation \& Regulatory Science},
  pages = {1--15},
  year = {2023},
  publisher = {Springer}
}

@article{doi:10.1177/1740774509356002,
author  =  {Beat Neuenschwander and Gorana Capkun-Niggli and Michael Branson and David J. Spiegelhalter},
title  = {Summarizing historical information on controls in clinical trials},
journal  =  {Clin Trials},
volume  =  {7},
number  =  {1},
pages  =  {5-18},
year  =  {2010},
doi  =  {10.1177/1740774509356002},
    note  = {PMID: 20156954},

URL  =  { 
        https://doi.org/10.1177/1740774509356002
    
}
}

@phdthesis{nikolakopoulos2016hybrid,
  title={Hybrid Bayesian-frequentist approaches for randomized trial design in small populations},
  author={Nikolakopoulos, Stavros},
  year={2016},
  school={Utrecht University}
}

@article{O’Hagan29032019,
author = {Anthony O’Hagan},
title = {Expert Knowledge Elicitation: Subjective but Scientific},
journal = {The American Statistician},
volume = {73},
number = {sup1},
pages = {69--81},
year = {2019},
publisher = {ASA Website} 
}

@book{robert2007bayesian,
  title = {The Bayesian choice: from decision-theoretic foundations to computational implementation},
  author = {Robert, Christian},
  year = {2007},
  publisher = {Springer Science \& Business Media}
}

@article{robertson2023response,
  title={Response-adaptive randomization in clinical trials: from myths to practical considerations},
  author={Robertson, David S and Lee, Kim May and L{\'o}pez-Kolkovska, Boryana C and Villar, Sof{\'\i}a S},
  journal={Statistical science: a review journal of the Institute of Mathematical Statistics},
  volume={38},
  number={2},
  pages={185},
  year={2023},
  publisher={Europe PMC Funders}
}

@article{rozeboom1960fallacy,
  title={The fallacy of the null-hypothesis significance test.},
  author={Rozeboom, William W},
  journal={Psychological bulletin},
  volume={57},
  number={5},
  pages={416},
  year={1960},
  publisher={American Psychological Association}
}

@article{rosner2020bayesian,
  title={Bayesian methods in regulatory science},
  author={Rosner, Gary L},
  journal={Statistics in biopharmaceutical research},
  volume={12},
  number={2},
  pages={130--136},
  year={2020},
  publisher={Taylor \& Francis}
}

@article{ruberg2023application,
  title = {Application of Bayesian approaches in drug development: starting a virtuous cycle},
  author = {Ruberg, Stephen J and Beckers, Francois and Hemmings, Rob and Honig, Peter and Irony, Telba and LaVange, Lisa and Lieberman, Grazyna and Mayne, James and Moscicki, Richard},
  journal = {Nat Rev Drug Discov},
  pages = {1--16},
  year = {2023},
  publisher = {Nature Publishing Group UK London}
}

@article{rufibach2016bayesian,
  title = {Bayesian predictive power: choice of prior and some recommendations for its use as probability of success in drug development},
  author = {Rufibach, Kaspar and Burger, Hans Ulrich and Abt, Markus},
  journal = {Pharm Stat},
  volume = {15},
  number = {5},
  pages = {438--446},
  year = {2016},
  publisher = {Wiley Online Library}
}

@article{sambucini2021bayesian,
  title = {Bayesian Sequential Monitoring of Single-Arm Trials: A Comparison of Futility Rules Based on Binary Data},
  author = {Sambucini, Valeria},
  journal = { Int J Environ Res Public Health},
  volume = {18},
  number = {16},
  pages = {8816},
  year = {2021},
  publisher = {MDPI}
}

@article{saville2014utility,
  title = {The utility of Bayesian predictive probabilities for interim monitoring of clinical trials},
  author = {Saville, Benjamin R and Connor, Jason T and Ayers, Gregory D and Alvarez, JoAnn},
  journal = {Clin Trials},
  volume = {11},
  number = {4},
  pages = {485--493},
  year = {2014},
  publisher = {SAGE Publications Sage UK: London, England}
}

@article{saville2023conditional,
  title = {Conditional Power: How Likely Is Trial Success?},
  author = {Saville, Benjamin R and Detry, Michelle A and Viele, Kert},
  journal = {JAMA},
  volume = {329},
  number = {6},
  pages = {508--509},
  year = {2023},
  publisher = {American Medical Association}
}

@article{shi2019control,
  title = {Control of type I error rates in Bayesian sequential designs},
  author = {Shi, Haolun and Yin, Guosheng and others},
  journal = {Bayesian Anal},
  volume = {14},
  number = {2},
  pages = {399--425},
  year = {2019},
  publisher = {International Society for Bayesian Analysis}
}

@article{simon1994some,
  title={Some practical aspects of the interim monitoring of clinical trials},
  author={Simon, Richard},
  journal={Statistics in medicine},
  volume={13},
  number={13-14},
  pages={1401--1409},
  year={1994},
  publisher={Wiley Online Library}
}

@article{SPIEGELHALTER19868,
title  =  "Monitoring clinical trials: Conditional or predictive power?",
journal  =  "Contr Clin Trials",
volume  =  "7",
number  =  "1",
pages  =  "8 - 17",
year  =  "1986",
issn  =  "0197-2456",
doi  =  "https://doi.org/10.1016/0197-2456(86)90003-6",
url  =  "http://www.sciencedirect.com/science/article/pii/0197245686900036",
author  =  "David J. Spiegelhalter and Laurence S. Freedman and Patrick R. Blackburn"
}

@article{spiegelhalter1994bayesian,
  title = {Bayesian approaches to randomized trials},
  author = {Spiegelhalter, David J. and Freedman, Laurence S. and Parmar, Mahesh K. B.},
  journal = { J Roy Stat Soc Series A
Stat Society},
  volume = {157},
  number = {3},
  pages = {357--387},
  year = {1994},
  publisher = {Wiley Online Library}
}

@book{spiegelhalter2004bayesian,
  title = {Bayesian approaches to clinical trials and health-care evaluation},
  author = {Spiegelhalter, David J. and Abrams, Keith R. and Myles, Jonathan P.},
  volume = {13},
  year = {2004},
  publisher = {John Wiley \& Sons}
}

@article{10.1214/088342304000000080,
author = {David J. Spiegelhalter},
title = {{Incorporating Bayesian Ideas into Health-Care Evaluation}},
volume = {19},
journal = {Statistical Science},
number = {1},
publisher = {Institute of Mathematical Statistics},
pages = {156 -- 174},
year = {2004},
doi = {10.1214/088342304000000080},
URL = {https://doi.org/10.1214/088342304000000080}
}

@article{storey2003positive,
  title = {The positive false discovery rate: a Bayesian interpretation and the q-value},
  author = {Storey, John D},
  journal = {Ann Stat},
  volume = {31},
  number = {6},
  pages = {2013--2035},
  year = {2003},
  publisher = {Institute of Mathematical Statistics}
}

@article{thall2019bayesian,
  title={Bayesian treatment comparison using parametric mixture priors computed from elicited histograms},
  author={Thall, Peter F and Ursino, Moreno and Baudouin, V{\'e}ronique and Alberti, Corinne and Zohar, Sarah},
  journal={Statistical methods in medical research},
  volume={28},
  number={2},
  pages={404--418},
  year={2019},
  publisher={SAGE Publications Sage UK: London, England}
}

@article{ventz2015bayesian,
  title = {Bayesian designs and the control of frequentist characteristics: a practical solution},
  author = {Ventz, Steffen and Trippa, Lorenzo},
  journal = {Biometrics},
  volume = {71},
  number = {1},
  pages = {218--226},
  year = {2015},
  publisher = {Wiley Online Library}
}

@article{ventz2017combining,
  title={Combining Bayesian experimental designs and frequentist data analyses: motivations and examples},
  author={Ventz, Steffen and Parmigiani, Giovanni and Trippa, Lorenzo},
  journal={Applied Stochastic Models in Business and Industry},
  volume={33},
  number={3},
  pages={302--313},
  year={2017},
  publisher={Wiley Online Library}
}

@article{wacholder2004assessing,
  title={Assessing the probability that a positive report is false: an approach for molecular epidemiology studies},
  author={Wacholder, Sholom and Chanock, Stephen and Garcia-Closas, Montserrat and El Ghormli, Laure and Rothman, Nathaniel},
  journal={Journal of the National Cancer Institute},
  volume={96},
  number={6},
  pages={434--442},
  year={2004},
  publisher={Oxford University Press}
}

@article{wang2019spy,
  title={I-SPY 2: a neoadjuvant adaptive clinical trial designed to improve outcomes in high-risk breast cancer},
  author={Wang, Haiyun and Yee, Douglas},
  journal={Current breast cancer reports},
  volume={11},
  number={4},
  pages={303--310},
  year={2019},
  publisher={Springer}
}

@article{doi:10.1080/00031305.2016.1154108,
author  =  {Ronald L. Wasserstein and Nicole A. Lazar},
title  =  {The ASA Statement on p-Values: Context, Process, and Purpose},
journal  =  {The American Statistician},
volume  =  {70},
number  =  {2},
pages  =  {129-133},
year   =  {2016},
publisher  =  {Taylor & Francis} 

}

@article{wathen2006implementation,
  title = {Implementation of backward induction for sequentially adaptive clinical trials},
  author = {Wathen, J Kyle and Christen, J Andr{\'e}s},
  journal = { J Comput Graph Stat},
  volume = {15},
  number = {2},
  pages = {398--413},
  year = {2006},
  publisher = {Taylor \& Francis}
}

@article{wiener2020methods,
  title = {Methods for clarifying criteria for study continuation at interim analysis},
  author = {Wiener, Laura E and Ivanova, Anastasia and Koch, Gary G},
  journal = {Pharm Stat},
  volume = {19},
  number = {5},
  pages = {720--732},
  year = {2020},
  publisher = {Wiley Online Library}
}

@article{yi2012hybridization,
  title = {Hybridization of conditional and predictive power for futility assessment in sequential clinical trials with time-to-event outcomes: A resampling approach},
  author = {Yi, Jing and Fang, Liang and Su, Zheng},
  journal = {Contemp Clin Trials},
  volume = {33},
  number = {1},
  pages = {138--142},
  year = {2012},
  publisher = {Elsevier}
}

@phdthesis{yu2019unified,
  title={Unified Approaches for Frequentist and Bayesian Methods in Two-Sample Clinical Trials with Binary Endpoints},
  author={Yu, Zhenning},
  year={2019},
school = {Medical University of South Carolina}
}

@article{Zaslavsky2012,
  doi  =  {10.1111/j.1541-0420.2012.01806.x},
  url  =  {https://doi.org/10.1111/j.1541-0420.2012.01806.x},
  year  =  {2012},
  month  =  {9},
  publisher  =  {Wiley},
  volume  =  {69},
  number  =  {1},
  pages  =  {157--163},
  author  =  {Boris G. Zaslavsky},
  title  =  {Bayesian Hypothesis Testing in Two-Arm Trials with Dichotomous Outcomes},
  journal  =  {Biometrics}
}

@article{https://doi.org/10.1002/sim.7685,
author  =  {Zhou, Ming and Tang, Qi and Lang, Lixin and Xing, Jun and Tatsuoka, Kay},
title  =  {Predictive probability methods for interim monitoring in clinical trials with longitudinal outcomes},
journal  =  {Stat Med},
volume  =  {37},
number  =  {14},
pages  =  {2187-2207},
doi  =  {https://doi.org/10.1002/sim.7685},
url  =  {https://onlinelibrary.wiley.com/doi/abs/10.1002/sim.7685},
year  =  {2018}
}

@article{zhou_ji_2023,
    author  =  {Tianjian Zhou and Yuan Ji},
    title  =  {On Bayesian Sequential Clinical Trial Designs},
    journal  =  {N Engl J Stat Data Sci},
    year  =  {2023},
    pages  =  {1--16},
    doi  =  {10.51387/23-NEJSDS24},
    issn  =  {2693-7166},
    publisher  =  {New England Statistical Society}
}

@article{doi:10.1177/0962280215595058,
author  =  {Han Zhu and Qingzhao Yu},
title  = {A Bayesian sequential design using alpha spending function to control type I error},
journal  =  {Stat Methods Med Res},
volume  =  {26},
number  =  {5},
pages  =  {2184-2196},
year  =  {2017},
doi  =  {10.1177/0962280215595058},
 URL  =  { 
        https://doi.org/10.1177/0962280215595058
    
}
}

\newpage 
\onecolumn

\begin{enumerate}

\item
\textit{Extended Stopping and Decision Rule }.  

\begin{algorithm}[H]
\SetAlgoLined

$\tau_e \gets \infty$ \;
$\tau_f\gets \infty $ \;
$\tau_{\varnothing} \gets \infty$ \;
$\tau \gets\infty$ \;
$n\gets 0$ \;
\While{$\tau = \infty$ {\bf and  } $\sigma_n\le N_{\max}$ }
{
$T_n \gets (\sigma_n+m)\wedge N_{\max}$\;
\uIf{ 
$\P_{\pi_{e}}(\bftheta _{1} - \bftheta _{0}  \leq \Delta  \vert  {\bf D}_{\sigma_n}) < \varepsilon_e $ }{
  $\tau_e \gets \sigma_n$  \;
  $\tau \gets \sigma_n$ \;  
  $\delta \gets d_e$ \;
  }
\uElseIf{ 
$\P_{\pi_{f}}(\bftheta _{1} - \bftheta _{0}  \geq 0  \vert  {\bf D}_{\sigma_n}) < \varepsilon_f$  }{
  $\tau_f \gets \sigma_n$ \;
  $\tau \gets \sigma_n$   \;
 $\delta \gets d_f$ \;  
}\ElseIf{
$\E_{\pi_f}\bigl( U_{ \bftau_e \wedge \bftau_f  \wedge T_n} ( 
 \delta( {\bf D}_{\bftau_e\wedge\bftau_e\wedge T_n}), 
 (\bftheta_0,\bftheta_1)) -(\bftau_e \wedge
 \bftau_f \wedge T_n-\sigma_n)\big\vert
 {\bf D}_{\sigma_n}\bigr)<0 $}{

 $\tau_{\varnothing} \gets \sigma_n$ \;
 $\tau \gets \sigma_n$ \;
 $\delta \gets d_{\varnothing} $
 \;
 }

$n \gets n+1$\;

}

\end{algorithm}

\end{enumerate}

\begin{appendices}
We provide analytic  formulae 
for updating sequentially the posterior distribution of 
$\{ \bftheta_0 > \bftheta_1\}$ in Bayesian filtering
with independent beta priors. However the 
the computation of Beta functions and hypergeometric functions appearing in these innovation
formulae for large values of the beta distributions parameters is prone to severe numerical instability. 
We describe an efficient Monte Carlo scheme
to evaluate these posterior probabilities.

\noappendicestocpagenum
\addappheadtotoc

\section{Notes on the analytic and numerical
computation
of some double Beta integrals \label{secA1}}

\begin{lemma}  

\begin{enumerate}
\item  A random variable $\bftheta$
is
$\mbox{Beta}( \alpha,\beta)$-distributed
if and only if 
$(1-\bftheta)$  
is $\mbox{Beta}( \beta ,\alpha)$-distributed
\item 
  The incomplete Beta function has representation
  \begin{align*} &
   \int_0^x y^{\alpha-1} (1-y)^{\beta-1} dy=\frac {x^{\alpha}}
   { \alpha }
   {}_2 F_1 \bigl ( \alpha , 1-\beta; \alpha+1; x \bigr) 
   = \frac {x^\alpha (1-x)^{\beta} }{ \alpha }  {}_2 F_1\bigl 
   ( \alpha +\beta, 1; \alpha+1; x \bigr) 
 & \\ &
 =\int_{1-x}^1 (1-y)^{\alpha-1} y^{\beta-1} dy
 = \frac{\Gamma(\alpha) \Gamma(\beta) }{\Gamma(\alpha+\beta) }
 -\frac{ (1-x)^{\beta} }{\beta } {}_2 F_1\bigl( \beta,
 1-\alpha,\beta+1, 1-x\bigr)  &\end{align*}
where ${}_2 F_1( a,b;c;x)$ is the Gauss hypergeometric function.
\item Gauss generalized hypergeometric function satisfies
  \begin{align*}&\frac{ \Gamma(\alpha+\beta) } { \Gamma(\alpha) \Gamma(\beta) }
   \int_0^1 x^{\alpha-1} (1-x)^{\beta-1} {}_p F_q( a_1,\dots, a_p; b_1,\dots,b_q; xy) dx
& \\ &   = 
   {}_{p+1} F_{q+1}( a_1,\dots, a_p,\alpha; b_1,\dots,b_q,\alpha+\beta; y)
 & \end{align*}
 \item  
 \begin{align*} &
  \int_0^1 x^{\gamma-1} (1-x)^{\rho-1}(1-z x)^{-\sigma}  {}_2 F_1\bigl( 
  \alpha, \beta; \gamma; x \bigr)dx
  & \\ &=(1-z)^{\sigma} \frac{  \Gamma( \gamma) \Gamma ( \rho) \Gamma( \gamma +\rho-\alpha-\beta) }
  { \Gamma( \gamma + \rho-\alpha) \Gamma( \gamma+ \rho-\beta) }
  {}_3 F_2\bigl(  \rho,\sigma,\gamma +\rho-\alpha-\beta; \gamma+\rho-\alpha, \gamma+
  \rho-\beta ; \frac{ z}{ z-1} \bigr)
 &\end{align*}
for $\Re(\gamma), \Re(\rho), \Re (\gamma+\rho-\alpha-\beta) > 0  , |\arg(1-x)| < \pi$ 
(7.512.9 in  \textcite{gradshteyn}).
 \item 
 \begin{align*} 
 (1-x)^{\alpha} ={}_1 F_0( \alpha; x) 
        = {}_2 F_1( \alpha,\beta; \beta;x)
       \end{align*}
 \end{enumerate}
\end{lemma}

\begin{lemma}
\begin{enumerate}
 \item
\begin{align*}
  & \P\bigl( \bftheta_1  < \bftheta_0)=
  \frac{ \Gamma( \alpha_0+\beta_0) \Gamma( \alpha_1+\beta_1)  }
  {     \Gamma( \alpha_0)   \Gamma(\beta_0) \Gamma( \alpha_1)   \Gamma( \beta_1)                   }
   \int_0^1  \int_0^{x}  
    y^{\alpha_1-1} (1-y)^{\beta_1-1}   dy \; x^{\alpha_0-1} (1-x)^{\beta_0-1} dx
=    & \\ &
\frac{ \Gamma( \alpha_0+\alpha_1) \Gamma( \beta_0 +\beta_1) \Gamma( \alpha_0+\beta_0) \Gamma( \alpha_1+\beta_1)  }
  {    \Gamma( \alpha_0+\alpha_1+\beta_0+\beta_1)  \Gamma( \alpha_0)   \Gamma(\beta_0) \Gamma( \alpha_1+1)   \Gamma( \beta_1)                    } \times & \\ &
 {}_3 F_2( \alpha_1 +\beta_1, 1, \alpha_0+\alpha_1; \alpha_1+1, \alpha_0+\alpha_1+\beta_0+\beta_1; 1)
 &\end{align*}
where the hypergeometric function ${}_3 F_2(a_1,a_2,a_3;b_1,b_2;z)$  is convergent for $|z|<1$, and also when
$|z|=1$ and $\Re\biggl(\sum_{k=1}^2 b_k - \sum_{j=1}^3 a_j \biggr) >0$.
\item We have also 
\begin{align*} &
\P( \bftheta_1 < c\bftheta_0)=
  \frac{ \Gamma( \alpha_0+\beta_0) \Gamma( \alpha_1+\beta_1)  }
  {     \Gamma( \alpha_0)   \Gamma(\beta_0) \Gamma( \alpha_1)   \Gamma( \beta_1)                   } 
   \int_0^1  \int_0^{x/c}  
    y^{\alpha_1-1} (1-y)^{\beta_1-1}   dy \; x^{\alpha_0-1} (1-x)^{\beta_0-1} dx
=     & \\ &
  \frac{ \Gamma( \alpha_0+\beta_0) \Gamma( \alpha_1+\beta_1) c^{\alpha_0} }
  {     \Gamma( \alpha_0)   \Gamma(\beta_0) \Gamma( \alpha_1)   \Gamma( \beta_1)                   }
   \int_0^1  \int_0^{u}  
    y^{\alpha_1-1} (1-y)^{\beta_1-1}   dy \; u^{\alpha_0-1} (1-cu)^{\beta_0-1} du    = 
    & \\ &
     \frac{ \Gamma(\alpha_0 +\beta_0 -1)
     \Gamma(\alpha_0+\alpha_1) \Gamma( \alpha_0+\beta_0) \Gamma( \alpha_1+\beta_1) c^{\alpha_0} (1-c)^{1-\beta_0 } }
  {\Gamma(\alpha_0 +\alpha_1+\beta_0 -1)   \Gamma( \alpha_0)^2   \Gamma(\beta_0) \Gamma(\alpha_1)\Gamma( \beta_1)}
 \times & \\ 
 & {}_3 F_2\biggl( 0, 1-\beta_0, \alpha_0 +\beta_1 -1; \alpha_0 +\alpha_1 + \beta_1 -1; \frac{c }{c-1} \biggr)
 & \end{align*}
\item   
  \begin{align*} & 
   \P( \bftheta_1 < \bftheta_0 + \Delta )= & \\ &
  \frac{ \Gamma( \alpha_0+\beta_0) \Gamma( \alpha_1+\beta_1)  }
  {     \Gamma( \alpha_0)   \Gamma(\beta_0) \Gamma( \alpha_1)   \Gamma( \beta_1)                   }
   \int_0^1  \int_0^{1\wedge( x+ \Delta) }  
    y^{\alpha_1-1} (1-y)^{\beta_1-1}   dy \; x^{\alpha_0-1} (1-x)^{\beta_0-1} dx 
    \end{align*}
    For $\Delta>0$
\begin{align*} & =
  \frac{ \Gamma( \alpha_0+\beta_0) \Gamma( \alpha_1+\beta_1) 
  \Delta^{\alpha_1} }
  {     \Gamma( \alpha_0)   \Gamma(\beta_0) \Gamma( \alpha_1)   \Gamma( \beta_1) \alpha_1 }
  \int_0^{1-\Delta}   \bigl(1+\frac{x }{\Delta} \bigr)^{\alpha_1}
{}_2 F_1\bigl( \alpha_1, 1-\beta_1; \alpha_1+1;
  x +\Delta \bigr)
    x^{\alpha_0-1} (1-x)^{\beta_0-1} dx   
    & \\ & +
   \frac{ \Gamma( \alpha_0+\beta_0)  }
  { \Gamma( \alpha_0) \Gamma(\beta_0) }   
    \int_{1-\Delta}^1 x^{\alpha_0-1} (1-x)^{\beta_0-1} dx
    = &\\ &  
\frac{ \Gamma( \alpha_0+\beta_0) \Gamma( \alpha_1+\beta_1) 
\Delta^{\alpha_1} }
  {     \Gamma( \alpha_0)   \Gamma(\beta_0) \Gamma( \alpha_1)   \Gamma( \beta_1) \alpha_1   }
  \int_0^{1-\Delta} 
{}_1 F_0\bigl(\alpha_1;
  -\frac {x} {\Delta} \bigr)
  {}_2 F_1\bigl( \alpha_1, 1-\beta_1; \alpha_1+1;
  x +\Delta \bigr)
    x^{\alpha_0-1} (1-x)^{\beta_0-1} dx   
    & \\ &
+ \frac{ \Gamma( \alpha_0+\beta_0)  }
  {     \Gamma( \alpha_0)   \Gamma(\beta_0)               }   
\frac{\Delta^{\beta_0}}{\beta_0}  {}_2 F_1(\beta_0, 1-\alpha_0;
  \beta_0+1;\Delta)
&  \end{align*}
\end{enumerate}\end{lemma}
\section{Innovation Gain formulae updating the posterior 
in Bayesian Filtering}\label{seca2}
For $\bftheta\sim\mbox{Beta}(\alpha,\beta)$, 
by using  integration by parts
we obtain the following
Stein equation for the Beta distribution
\begin{align*} &  \frac{\alpha \beta}{ (\alpha +\beta)(\alpha+\beta+1)}
\E_{\alpha+1,\beta+1} \bigl[ \partial f(\bftheta) \bigr ]  =
 \E_{\alpha,\beta}\bigl[ \bftheta (1-\bftheta) \partial f( \bftheta) \bigr) = & \\ & \E_{\alpha,\beta}\bigl[ \bigl( (\beta+\alpha) \bftheta
 -\alpha \bigr) f(\bftheta) \bigr]
 = \alpha \bigl\{
 \E_{\alpha+1,\beta}\bigl[ f(\bftheta) \bigr] -  \E_{\alpha,\beta}\bigl[ f(\bftheta)  \bigr]  \bigr\}
&\end{align*}
For $f(\theta)={\bf 1}(\theta > t)$, $\partial f(\theta)=\Delta_{t}(\theta)$, the Dirac delta function, which gives
\begin{align*} &
\P_{\alpha+1,\beta}( \bftheta > t ) =
\P_{\alpha,\beta}( \bftheta > t )  + \frac { \Gamma( \alpha+\beta)  }{ \Gamma(\alpha+1 ) \Gamma(\beta) } t^{\alpha} (1-t)^{\beta}
& \\ &
= \P_{\alpha,\beta}( \bftheta > t )  +  \alpha^{-1} \mbox{Beta}( \alpha,\beta)^{-1} t^{\alpha} (1-t)^{\beta} &
\end{align*}
which can be used to update the posterior recursively. We have also
\begin{align*}
\P_{\alpha+1,\beta}( \bftheta \le t ) -
\P_{\alpha,\beta}( \bftheta \le  t )  = -  
 \alpha^{-1} \mbox{Beta}( \alpha,\beta)^{-1}
t^{\alpha} (1-t)^{\beta}
\end{align*}
and
\begin{align*} &
 \P_{\alpha,\beta+1}( \bftheta \le t )= \P_{\alpha,\beta+1}( 1-\bftheta \ge 1- t )= 
 \P_{\beta+1,\alpha}( \bftheta \ge 1- t )= & \\ &  \P_{\alpha,\beta}(\bftheta \le  t ) +  \beta^{-1} \mbox{Beta}( \alpha,\beta)^{-1} t^{\alpha} (1-t)^{\beta}
&\end{align*}
By integrating we obtain for $\bftheta_0,\bftheta_1$ with independent  Beta$(\alpha_i,\beta_i)$ priors 
\begin{align*} &
 \bigl( \P_{\alpha_0+1,\beta_0}\otimes \P_{\alpha_1,\beta_1} \bigr)\bigl( \bftheta_0 > \bftheta_1 \bigr) -
 \bigl( \P_{\alpha_0,\beta_0}\otimes \P_{\alpha_1,\beta_1} \bigr)\bigl( \bftheta_0 > \bftheta_1 \bigr) = & \\ &\frac { \Gamma( \alpha_0+\beta_0)  }{ \Gamma(\alpha_0+1 ) \Gamma(\beta_0) } \frac { \Gamma( \alpha_1+\beta_1)  }{ \Gamma(\alpha_1) \Gamma(\beta_1) }
 \frac{\Gamma( \alpha_0+ \alpha_1)  \Gamma( \beta_0 +\beta_1)  } {\Gamma( \alpha_0+ \alpha_1 + \beta_0 +\beta_1)  }
 & \\ &
 =  \frac{ \mbox{Beta}(\alpha_0+\alpha_1,\beta_0 +\beta_1 )  }{ \alpha_0 \mbox{Beta}( \alpha_0,\beta_0) \mbox{Beta}(\alpha_1,\beta_1) } &
\end{align*}  and
\begin{align*} &
 \bigl( \P_{\alpha_0,\beta_0+1}\otimes \P_{\alpha_1,\beta_1} \bigr)\bigl( \bftheta_0 \le \bftheta_1 \bigr) -
 \bigl( \P_{\alpha_0,\beta_0}\otimes \P_{\alpha_1,\beta_1} \bigr)\bigl( \bftheta_0 \le \bftheta_1 \bigr)
 & \\ &
 =  \frac{ \mbox{Beta}(\beta_0 +\beta_1,\alpha_0+\alpha_1 )  }{ \beta_0 
 \mbox{Beta}( \beta_0,\alpha_0) \mbox{Beta}( \beta_1, \alpha_1) } &
\end{align*}
see \textcite{Zaslavsky2012}.
We derive also expressions
for the innovation gain in
the Bayes filtering formula
sequentially updating the posterior
distribution
\begin{align*} &
 \bigl( \P_{\alpha_0+1  ,\beta_0} \otimes  \P_{\alpha_1 , \beta_1} \bigr) \bigl( \bftheta_0 > \bftheta_1 + {\Delta}  \bigr)
-\bigl( \P_{\alpha_0,  \beta_0} \otimes  \P_{\alpha_1 , \beta_1} \bigr) \bigl( \bftheta_0 > \bftheta_1 + {\Delta}  \bigr) = & \\ &
 \alpha_0^{-1} \mbox{Beta}( \alpha_0,\beta_0)^{-1} \mbox{Beta}( \alpha_1,\beta_1)^{-1}
 \int_{{\Delta}^-}^{1-{\Delta}^+}  t^{\alpha_1 -1} (1-t)^{\beta_1-1} (t+{\Delta})^{\alpha_0} (1-{\Delta}-t)^{\beta_0} dt &
 \\ & = \alpha_0^{-1} \mbox{Beta}( \alpha_0,\beta_0)^{-1} \E_{\alpha_1,\beta_1} \biggl(  \bigl\{(\bftheta_1+{\Delta})^+\bigr\}^{\alpha_0}
\bigl\{  (1-\bftheta_1-{\Delta})^+ \bigr\}^{\beta_0} \biggr) 
= & \\ & 
\frac{\beta_0}{ (\alpha_0+\beta_0)(\alpha_0+\beta_0+1) \mbox{Beta}(\alpha_1,\beta_1)}
\E_{\alpha_0+1,\beta_0+1}\biggl (  \bigl\{( \bftheta_0-{\Delta})^+\bigr\}^{\alpha_1-1}
\bigl\{ ( 1+{\Delta} - \bftheta_0)^+ \bigr\}^{\beta_1-1} \biggr) =
& \\ & \alpha_0^{-1} \mbox{Beta}( \alpha_0,\beta_0)^{-1}\mbox{Beta}( \alpha_1,\beta_1)^{-1} \times  & \\ &\sum_{k=0}^{\infty} \sum_{\ell=0}^{k} {\Delta}^{k} (-1)^{\ell} \frac{ ( \beta_0)_\ell ( \alpha_0)_{k-\ell} }{ \ell!(k-\ell)!}
\int_{{\Delta}^-}^{1-{\Delta}^+}  (1- t)^{\beta_0+ \beta_1 -\ell -1 }t^{\alpha_0+ \alpha_1 + \ell -k -1 } dt & \\ &
= \alpha_0^{-1} \mbox{Beta}( \alpha_0,\beta_0)^{-1}\mbox{Beta}( \alpha_1,\beta_1)^{-1} \times  & \\ &
\sum_{k=0}^{\infty} \sum_{\ell=0}^{k} {\Delta}^{k} (-1)^{\ell}  \frac{  ( \beta_0)_\ell( \alpha_0)_{k-\ell} }
{ \ell!(k-\ell)!( \alpha_0+ \alpha_1+\ell-k )} \times
& \\ & 
\biggl\{
(1-{\Delta}^+)^{\alpha_0+ \alpha_1+\ell-k} 
{}_2 F_1(  \alpha_0 +\alpha_1 +\ell- k, 1+\ell  -\beta_0-\beta_1;1 + \alpha_0 + \alpha_1 + \ell -k;1-{\Delta}^+) 
& \\ &  - ({\Delta}^-)^{\alpha_0+ \alpha_1+\ell-k}  
{}_2 F_1(  \alpha_0 +\alpha_1 +\ell- k,1+ \ell  -\beta_0-\beta_1;1 +\alpha_0 + \alpha_1 + \ell -k;{\Delta}^-) 
\biggr\} &
\end{align*}
with ${\Delta}^+=\max\{ {\Delta},0\}$ ${\Delta}^-= \max\{-{\Delta},0\}$, 
and we have used the generalized Newton binomial formula where $(\alpha)_k=\alpha (\alpha-1) \dots (\alpha-k+1)$
is the Pochammer symbol.
We have also the updates
\begin{align*} & 
 \bigl( \P_{\alpha_0+1  ,\beta_0} \otimes  \P_{\alpha_1 , \beta_1} \bigr) \bigl( \bftheta_0 +{\Delta} < \bftheta_1  \bigr)
-\bigl( \P_{\alpha_0,  \beta_0} \otimes  \P_{\alpha_1 , \beta_1} \bigr) \bigl( \bftheta_0 +{\Delta} <\bftheta_1   \bigr) = &
 \\ & = - \alpha_0^{-1} \mbox{Beta}( \alpha_0,\beta_0)^{-1} \E_{\alpha_1,\beta_1} \biggl(  \bigl\{(\bftheta_1-{\Delta})^+\bigr\}^{\alpha_0}
\bigl\{  (1-\bftheta_1+{\Delta})^+ \bigr\}^{\beta_0} \biggr) 
&
\\ &
 \bigl( \P_{\alpha_0  ,\beta_0+1} \otimes  \P_{\alpha_1 , \beta_1} \bigr) \bigl( \bftheta_0 > \bftheta_1 + {\Delta}  \bigr)
-\bigl( \P_{\alpha_0,  \beta_0} \otimes  \P_{\alpha_1 , \beta_1} \bigr) \bigl( \bftheta_0 > \bftheta_1 + {\Delta}  \bigr)
= & \\ &
 \bigl( \P_{\beta_0+1,\alpha_0} \otimes  \P_{\beta_1,\alpha_1 } \bigr) \bigl( \bftheta_1 > \bftheta_0 + {\Delta}  \bigr)
-\bigl( \P_{\beta_0,\alpha_0} \otimes  \P_{\beta_1,\alpha_1} \bigr) \bigl( \bftheta_1> \bftheta_0 + {\Delta}  \bigr) = & \\ &
-\beta_0^{-1} \mbox{Beta}(\beta_0, \alpha_0)^{-1} \E_{\beta_1,\alpha_1} \biggl(  \bigl\{(\bftheta_1-{\Delta})^+\bigr\}^{\beta_0}
\bigl\{  (1-\bftheta_1+{\Delta})^+ \bigr\}^{\alpha_0} \biggr) 
& \\ &
\bigl( \P_{\alpha_0  ,\beta_0+1} \otimes  \P_{\alpha_1 , \beta_1} \bigr) \bigl( \bftheta_0 +{\Delta} < \bftheta_1  \bigr)
-\bigl( \P_{\alpha_0,  \beta_0} \otimes  \P_{\alpha_1 , \beta_1} \bigr) \bigl( \bftheta_0 +{\Delta} <\bftheta_1   \bigr)& \\ &
= \beta_0^{-1} \mbox{Beta}(\beta_0, \alpha_0)^{-1} \E_{\beta_1,\alpha_1} \biggl(  \bigl\{(\bftheta_1+{\Delta})^+\bigr\}^{\beta_0}
\bigl\{  (1-\bftheta_1-{\Delta})^+ \bigr\}^{\alpha_0} \biggr)
&
\end{align*}
When $\alpha_j,\beta_j\in \N$,  $\forall j=0,1\dots,K$,
the innovation gain in the filtering formula for
$\P( \bftheta_0+\Delta < \bftheta_1 | {\mathcal D}_t)$
has  analytic expression.

\section{ Efficient Monte Carlo approximation}\label{seca3}
The computation of 
 Beta functions and hypergeometric
 functions appearing in the
 innovation formulae
  for large values of the beta distribution
  parameters is prone to severe numerical
 instability. A simple and  robust numerical alternative to   evaluate the 
 posterior probabilities
 $\P( \bftheta_0 > \bftheta_1 | {\mathcal D}_t) $  and
$\P( \bftheta_1 > \bftheta_0 +\Delta | {\mathcal D}_t) $ is by using plain
Monte Carlo in the most
efficient way.
Let $X,Y$ be independent random
variables with respective cumulative
distribution functions $F,G$.
An estimator of 
\begin{align*} \P(  X>Y ) = 
\int_{\R} \int_{\R} 
{\bf 1}( x>y) F(dx)G(dy),
\end{align*} based on independent realizations  $(X_k,Y_k:
k=1,\dots,n),$
is
given  by
\begin{align*}
\widehat\P_n(X >Y )
= \int_{\R} \int_{\R} 
{\bf 1}( x> y) F_n(dx)G_n(dy)
=\frac{1}{n^2}
\sum_{j=1}^n
\sum_{k=1}^n
{\bf 1}( X_j > Y_k),
\end{align*}
where $F_n, G_n$
are the respective
empirical processes.
Asymptotically
$\sqrt n (F_n(x)-F(x) )\longrightarrow B(x)$
and
$\sqrt n ( G_n(y)-G(y))\longrightarrow 
Z(y)$,
which are
zero mean Brownian
bridge processes
$B(x),Z(y)$
with respective
covariances
$\E(B(x)B(x^{'})= F(x\wedge x^{'})-F(x)F(x^{'})$
and 
$\E(Z(y)Z(y^{'}))= G(y\wedge y^{'})-G(y)G(y^{'})$.
The estimator
$\widehat \P_n( X>Y)$
is unbiased 
and, by the functional
delta method,
\begin{align*}
\sqrt n \bigl( 
\widehat\P_n(X >Y) )
-  \P( X >Y)
\bigr)
\end{align*}
is asymptotically
zero mean Gaussian
with variance
\begin{align*}& \sigma^2
=\int_{\R} (1-F(y))^2 G(dy)
-\biggl(
\int_{\R} 
(1-F(y)) G(dy) \biggr)^2
+ \int_{\R} G(x)^2 F(dx)
-\biggl(
\int_{\R} 
G(x) F(dx) \biggr)^2
&  \\ &
= \text{Var}_G\bigl(
F(Y)\bigr)
+ \text{Var}_F\bigl(
G(X) \bigr).
\end{align*}
Computing
$\widehat P_n$
requires $n$ independent
samples from both distributions $F$ and $G$,
sorting the combined samples and making on average
$2n \log(2n)$
comparisons, achieving
asymptotic standard deviation $\sigma n^{-1/2}$.
The naive estimator based on the same
samples
\begin{align*}
\check \P_n(X > Y)
= \frac{1}{n}
\sum_{k=1}^n {\bf 1}
(X_k > Y_k),
\end{align*}
requiring $n$ comparisons,
is also unbiased and
\begin{align*}
\sqrt{n}\bigl(\check \P_n(X > Y)-
\P(X > Y) \bigr)
\end{align*}
is asymptotically
Gaussian with
zero mean
and variance
\begin{align*} &
\eta^2=
\P\bigl( X > Y)
- 
\P\bigl(X > Y) ^2
= \E_{F}(G(X))-\E_{F}(G(X))^2
= \E_{G}(1-F(Y))-\E_{G}(1-F(Y))^2.
&
\end{align*}
For $X,X^{'},Y,Y^{'}$
independent random variables with
$X,X^{'}\sim F$
and $Y,Y^{'}\sim G$,
\begin{align*}
\eta^2-\sigma^2 = \frac{1} {4} \E\biggl[\biggl(
{\bf 1}(X > Y)- {\bf 1}(X>Y^{'})+{\bf 1}(X^{'}> Y^{'})
- {\bf 1} (X^{'}>Y) \biggr) ^2 \biggr]\ge 0.
\end{align*}
In practice the computational cost of comparing variables is much smaller than the cost of  
sampling random variables, and it is computationally more efficient to  make $2n\log (2n) >n$ comparisons   in order to achieve  the  smaller constant $\sigma^2\le\eta^2$  in the asymptotic  error variance (see
also \textcite{SangitaIsha2023}).

\end{appendices}

\end{document}